
%
%
%
%
\def\doublespace{\baselineskip=20pt plus 2pt\lineskip=3pt minus 
     1pt\lineskiplimit=2pt}

\def\np{\vfill\eject}
\def\singlespace{\normalbaselines}


\parindent=20pt

%
%
\expandafter\ifx\csname phyzzx\endcsname\relax
 \message{It is better to use PHYZZX format than to
          \string\input\space PHYZZX}\else
 \wlog{PHYZZX macros are already loaded and are not
          \string\input\space again}%
 \endinput \fi
\catcode`\@=11 
\let\rel@x=\relax
\let\n@expand=\relax
\def\pr@tect{\let\n@expand=\noexpand}
\let\protect=\pr@tect
\let\gl@bal=\global
%
%
%
\newfam\cpfam
\newdimen\b@gheight             \b@gheight=12pt
\newcount\f@ntkey               \f@ntkey=0
\def\f@m{\afterassignment\samef@nt\f@ntkey=}
\def\samef@nt{\fam=\f@ntkey \the\textfont\f@ntkey\rel@x}
\def\setstr@t{\setbox\strutbox=\hbox{\vrule height 0.85\b@gheight
                                depth 0.35\b@gheight width\z@ }}
%
%
%
%
%

\font\fourteenrm  =cmr10 scaled\magstep2
\font\twelverm    =cmr12
\font\ninerm      =cmr9
\font\sixrm       =cmr6

\font\fourteenbf  =cmbx10 scaled\magstep2
\font\twelvebf    =cmbx12
\font\ninebf      =cmbx9
\font\sixbf       =cmbx6
\font\seventeeni  =cmmi10 scaled\magstep3    \skewchar\seventeeni='177
\font\fourteeni   =cmmi10 scaled\magstep2     \skewchar\fourteeni='177
\font\twelvei     =cmmi12                       \skewchar\twelvei='177
\font\ninei       =cmmi9                          \skewchar\ninei='177
\font\sixi        =cmmi6                           \skewchar\sixi='177
\font\seventeensy =cmsy10 scaled\magstep3    \skewchar\seventeensy='60
\font\fourteensy  =cmsy10 scaled\magstep2     \skewchar\fourteensy='60
\font\twelvesy    =cmsy10 scaled\magstep1       \skewchar\twelvesy='60
\font\ninesy      =cmsy9                          \skewchar\ninesy='60
\font\sixsy       =cmsy6                           \skewchar\sixsy='60

\font\fourteenex  =cmex10 scaled\magstep2
\font\twelveex    =cmex10 scaled\magstep1

\font\fourteensl  =cmsl10 scaled\magstep2
\font\twelvesl    =cmsl12
\font\ninesl      =cmsl9

\font\fourteenit  =cmti10 scaled\magstep2
\font\twelveit    =cmti12
\font\nineit      =cmti9
\font\fourteentt  =cmtt10 scaled\magstep2
\font\twelvett    =cmtt12
\font\fourteencp  =cmcsc10 scaled\magstep2
\font\twelvecp    =cmcsc10 scaled\magstep1
\font\tencp       =cmcsc10
%
%
\def\fourteenf@nts{\relax
    \textfont0=\fourteenrm          \scriptfont0=\tenrm
      \scriptscriptfont0=\sevenrm
    \textfont1=\fourteeni           \scriptfont1=\teni
      \scriptscriptfont1=\seveni
    \textfont2=\fourteensy          \scriptfont2=\tensy
      \scriptscriptfont2=\sevensy
    \textfont3=\fourteenex          \scriptfont3=\twelveex
      \scriptscriptfont3=\tenex
    \textfont\itfam=\fourteenit     \scriptfont\itfam=\tenit
    \textfont\slfam=\fourteensl     \scriptfont\slfam=\tensl
    \textfont\bffam=\fourteenbf     \scriptfont\bffam=\tenbf
      \scriptscriptfont\bffam=\sevenbf
    \textfont\ttfam=\fourteentt
    \textfont\cpfam=\fourteencp }
\def\twelvef@nts{\relax
    \textfont0=\twelverm          \scriptfont0=\ninerm
      \scriptscriptfont0=\sixrm
    \textfont1=\twelvei           \scriptfont1=\ninei
      \scriptscriptfont1=\sixi
    \textfont2=\twelvesy           \scriptfont2=\ninesy
      \scriptscriptfont2=\sixsy
    \textfont3=\twelveex          \scriptfont3=\tenex
      \scriptscriptfont3=\tenex
    \textfont\itfam=\twelveit     \scriptfont\itfam=\nineit
    \textfont\slfam=\twelvesl     \scriptfont\slfam=\ninesl
    \textfont\bffam=\twelvebf     \scriptfont\bffam=\ninebf
      \scriptscriptfont\bffam=\sixbf
    \textfont\ttfam=\twelvett
    \textfont\cpfam=\twelvecp }
\def\tenf@nts{\relax
    \textfont0=\tenrm          \scriptfont0=\sevenrm
      \scriptscriptfont0=\fiverm
    \textfont1=\teni           \scriptfont1=\seveni
      \scriptscriptfont1=\fivei
    \textfont2=\tensy          \scriptfont2=\sevensy
      \scriptscriptfont2=\fivesy
    \textfont3=\tenex          \scriptfont3=\tenex
      \scriptscriptfont3=\tenex
    \textfont\itfam=\tenit     \scriptfont\itfam=\seveni  
    \textfont\slfam=\tensl     \scriptfont\slfam=\sevenrm 
    \textfont\bffam=\tenbf     \scriptfont\bffam=\sevenbf
      \scriptscriptfont\bffam=\fivebf
    \textfont\ttfam=\tentt
    \textfont\cpfam=\tencp }
%
%

%
\def\rm{\n@expand\f@m0 }
\def\mit{\n@expand\f@m1 }         
\def\cal{\n@expand\f@m2 }
\def\it{\n@expand\f@m\itfam}
\def\sl{\n@expand\f@m\slfam}
\def\bf{\n@expand\f@m\bffam}
\def\tt{\n@expand\f@m\ttfam}
\def\caps{\n@expand\f@m\cpfam}    
\def\em@{\rel@x\ifnum\f@ntkey=0 \it \else
        \ifnum\f@ntkey=\bffam \it \else \rm \fi \fi }
\def\em{\n@expand\em@}
\def\fourteenpoint{\fourteenf@nts \samef@nt \b@gheight=14pt \setstr@t }
\def\twelvepoint{\twelvef@nts \samef@nt \b@gheight=12pt \setstr@t }
\def\tenpoint{\tenf@nts \samef@nt \b@gheight=10pt \setstr@t }
\normalbaselineskip = 19.2pt plus 0.2pt minus 0.1pt 
\normallineskip = 1.5pt plus 0.1pt minus 0.1pt
\normallineskiplimit = 1.5pt
\newskip\normaldisplayskip
\normaldisplayskip = 14.4pt plus 3.6pt minus 10.0pt 
\newskip\normaldispshortskip
\normaldispshortskip = 6pt plus 5pt
\newskip\normalparskip
\normalparskip = 6pt plus 2pt minus 1pt
\newskip\skipregister
\skipregister = 5pt plus 2pt minus 1.5pt
\newif\ifsingl@
\newif\ifdoubl@
\newif\iftwelv@  \twelv@true
\def\singlespace{\singl@true\doubl@false\spaces@t}
\def\doublespace{\singl@false\doubl@true\spaces@t}
\def\normalspace{\singl@false\doubl@false\spaces@t}
\def\Tenpoint{\tenpoint\twelv@false\spaces@t}
\def\Twelvepoint{\twelvepoint\twelv@true\spaces@t}
\def\spaces@t{\rel@x
      \iftwelv@ \ifsingl@\subspaces@t3:4;\else\subspaces@t1:1;\fi
       \else \ifsingl@\subspaces@t3:5;\else\subspaces@t4:5;\fi \fi
      \ifdoubl@ \multiply\baselineskip by 5
         \divide\baselineskip by 4 \fi }
\def\subspaces@t#1:#2;{
      \baselineskip = \normalbaselineskip
      \multiply\baselineskip by #1 \divide\baselineskip by #2
      \lineskip = \normallineskip
      \multiply\lineskip by #1 \divide\lineskip by #2
      \lineskiplimit = \normallineskiplimit
      \multiply\lineskiplimit by #1 \divide\lineskiplimit by #2
      \parskip = \normalparskip
      \multiply\parskip by #1 \divide\parskip by #2
      \abovedisplayskip = \normaldisplayskip
      \multiply\abovedisplayskip by #1 \divide\abovedisplayskip by #2
      \belowdisplayskip = \abovedisplayskip
      \abovedisplayshortskip = \normaldispshortskip
      \multiply\abovedisplayshortskip by #1
        \divide\abovedisplayshortskip by #2
      \belowdisplayshortskip = \abovedisplayshortskip
      \advance\belowdisplayshortskip by \belowdisplayskip
      \divide\belowdisplayshortskip by 2
      \smallskipamount = \skipregister
      \multiply\smallskipamount by #1 \divide\smallskipamount by #2
      \medskipamount = \smallskipamount \multiply\medskipamount by 2
      \bigskipamount = \smallskipamount \multiply\bigskipamount by 4 }
\def\normalbaselines{ \baselineskip=\normalbaselineskip
   \lineskip=\normallineskip \lineskiplimit=\normallineskip
   \iftwelv@\else \multiply\baselineskip by 4 \divide\baselineskip by 5
     \multiply\lineskiplimit by 4 \divide\lineskiplimit by 5
     \multiply\lineskip by 4 \divide\lineskip by 5 \fi }
\Twelvepoint  
\interlinepenalty=50
\interfootnotelinepenalty=5000
\predisplaypenalty=9000
\postdisplaypenalty=500
\hfuzz=1pt
\vfuzz=0.2pt
\newdimen\HOFFSET  \HOFFSET=0pt
\newdimen\VOFFSET  \VOFFSET=0pt
\newdimen\HSWING   \HSWING=0pt
\dimen\footins=8in
%
%
%
\newskip\pagebottomfiller
\pagebottomfiller=\z@ plus \z@ minus \z@
\def\pagecontents{
   \ifvoid\topins\else\unvbox\topins\vskip\skip\topins\fi
   \dimen@ = \dp255 \unvbox255
   \vskip\pagebottomfiller
   \ifvoid\footins\else\vskip\skip\footins\footrule\unvbox\footins\fi
   \ifr@ggedbottom \kern-\dimen@ \vfil \fi }
\def\makeheadline{\vbox to 0pt{ \skip@=\topskip
      \advance\skip@ by -12pt \advance\skip@ by -2\normalbaselineskip
      \vskip\skip@ \line{\vbox to 12pt{}\the\headline} \vss
      }\nointerlineskip}
\def\makefootline{\baselineskip = 1.5\normalbaselineskip
                 \line{\the\footline}}
\newif\iffrontpage
\newif\ifp@genum
\def\nopagenumbers{\p@genumfalse}
\def\pagenumbers{\p@genumtrue}
\pagenumbers
\newtoks\paperheadline
\newtoks\paperfootline
\newtoks\letterheadline
\newtoks\letterfootline
\newtoks\letterinfo
\newtoks\date
\paperheadline={\hfil}
\paperfootline={\hss\iffrontpage\else\ifp@genum\tenrm\folio\hss\fi\fi}
\letterheadline{\iffrontpage \hfil \else
    \rm \ifp@genum page~~\folio\fi \hfil\the\date \fi}
\letterfootline={\iffrontpage\the\letterinfo\else\hfil\fi}
\letterinfo={\hfil}
\def\monthname{\rel@x\ifcase\month 0/\or January\or February\or
   March\or April\or May\or June\or July\or August\or September\or
   October\or November\or December\else\number\month/\fi}
\def\today{\monthname~\number\day, \number\year}
\date={\today}
\headline=\paperheadline 
\footline=\paperfootline 
\countdef\pageno=1      \countdef\pagen@=0
\countdef\pagenumber=1  \pagenumber=1
\def\advancepageno{\gl@bal\advance\pagen@ by 1
   \ifnum\pagenumber<0 \gl@bal\advance\pagenumber by -1
    \else\gl@bal\advance\pagenumber by 1 \fi
    \gl@bal\frontpagefalse  \swing@ }
\def\folio{\ifnum\pagenumber<0 \romannumeral-\pagenumber
           \else \number\pagenumber \fi }
\def\swing@{\ifodd\pagenumber \gl@bal\advance\hoffset by -\HSWING
             \else \gl@bal\advance\hoffset by \HSWING \fi }
\def\footrule{\dimen@=\prevdepth\nointerlineskip
   \vbox to 0pt{\vskip -0.25\baselineskip \hrule width 0.35\hsize \vss}
   \prevdepth=\dimen@ }
\let\footnotespecial=\rel@x
\newdimen\footindent
\footindent=24pt
\def\Textindent#1{\noindent\llap{#1\enspace}\ignorespaces}
\def\Vfootnote#1{\insert\footins\bgroup
   \interlinepenalty=\interfootnotelinepenalty \floatingpenalty=20000
   \singl@true\doubl@false\Tenpoint
   \splittopskip=\ht\strutbox \boxmaxdepth=\dp\strutbox
   \leftskip=\footindent \rightskip=\z@skip
   \parindent=0.5\footindent \parfillskip=0pt plus 1fil
   \spaceskip=\z@skip \xspaceskip=\z@skip \footnotespecial
   \Textindent{#1}\footstrut\futurelet\next\fo@t}

\def\vfootnote#1{\Vfootnote{${#1}$}}
\def\footnote#1{\attach{#1}\vfootnote{#1}}

\let\footsymbol=\star
\newcount\lastf@@t           \lastf@@t=-1
\newcount\footsymbolcount    \footsymbolcount=0
\newif\ifPhysRev
\def\bumpfootsymbolcount{\rel@x
   \iffrontpage \bumpfootsymbolpos \else \advance\lastf@@t by 1
     \ifPhysRev \bumpfootsymbolneg \else \bumpfootsymbolpos \fi \fi
   \gl@bal\lastf@@t=\pagen@ }
\def\bumpfootsymbolpos{\ifnum\footsymbolcount <0
                            \gl@bal\footsymbolcount =0 \fi
    \ifnum\lastf@@t<\pagen@ \gl@bal\footsymbolcount=0
     \else \gl@bal\advance\footsymbolcount by 1 \fi }
\def\bumpfootsymbolneg{\ifnum\footsymbolcount >0
             \gl@bal\footsymbolcount =0 \fi
         \gl@bal\advance\footsymbolcount by -1 }
\def\fd@f#1 {\xdef\footsymbol{\mathchar"#1 }}
\def\generatefootsymbol{\ifcase\footsymbolcount \fd@f 13F \or \fd@f 279
        \or \fd@f 27A \or \fd@f 278 \or \fd@f 27B \else
        \ifnum\footsymbolcount <0 \fd@f{023 \number-\footsymbolcount }
         \else \fd@f 203 {\loop \ifnum\footsymbolcount >5
                \fd@f{203 \footsymbol } \advance\footsymbolcount by -1
                \repeat }\fi \fi }

\def\nonfrenchspacing{\sfcode`\.=3001 \sfcode`\!=3000 \sfcode`\?=3000
        \sfcode`\:=2000 \sfcode`\;=1500 \sfcode`\,=1251 }
\nonfrenchspacing
\newdimen\d@twidth
{\setbox0=\hbox{s.} \gl@bal\d@twidth=\wd0 \setbox0=\hbox{s}
        \gl@bal\advance\d@twidth by -\wd0 }
\def\removehglue{\loop \unskip \ifdim\lastskip >\z@ \repeat }
\def\roll@ver#1{\removehglue \nobreak \count255 =\spacefactor \dimen@=\z@
        \ifnum\count255 =3001 \dimen@=\d@twidth \fi
        \ifnum\count255 =1251 \dimen@=\d@twidth \fi
    \iftwelv@ \kern-\dimen@ \else \kern-0.83\dimen@ \fi
   #1\spacefactor=\count255 }
\def\step@ver#1{\rel@x \ifmmode #1\else \ifhmode
        \roll@ver{${}#1$}\else {\setbox0=\hbox{${}#1$}}\fi\fi }
\def\attach#1{\step@ver{\strut^{\mkern 2mu #1} }}
%
%
%
\newcount\chapternumber      \chapternumber=0
\newcount\sectionnumber      \sectionnumber=0
\newcount\equanumber         \equanumber=0
\let\chapterlabel=\rel@x
\let\sectionlabel=\rel@x
\newtoks\chapterstyle        \chapterstyle={\Number}
\newtoks\sectionstyle        \sectionstyle={\chapterlabel.\Number}
\newskip\chapterskip         \chapterskip=\bigskipamount
\newskip\sectionskip         \sectionskip=\medskipamount
\newskip\headskip            \headskip=8pt plus 3pt minus 3pt
\newdimen\chapterminspace    \chapterminspace=15pc
\newdimen\sectionminspace    \sectionminspace=10pc
\newdimen\referenceminspace  \referenceminspace=20pc
\def\chapterreset{\gl@bal\advance\chapternumber by 1
   \ifnum\equanumber<0 \else\gl@bal\equanumber=0\fi
   \sectionnumber=0 \let\sectionlabel=\rel@x
   {\pr@tect\xdef\chapterlabel{\the\chapterstyle{\the\chapternumber}}}}
\def\alphabetic#1{\count255='140 \advance\count255 by #1\char\count255}
\def\Alphabetic#1{\count255='100 \advance\count255 by #1\char\count255}
\def\Roman#1{\uppercase\expandafter{\romannumeral #1}}
\def\roman#1{\romannumeral #1}
\def\Number#1{\number #1}
\def\BLANC#1{}
\def\titleparagraphs{\interlinepenalty=9999
     \leftskip=0.03\hsize plus 0.22\hsize minus 0.03\hsize
     \rightskip=\leftskip \parfillskip=0pt
     \hyphenpenalty=9000 \exhyphenpenalty=9000
     \tolerance=9999 \pretolerance=9000
     \spaceskip=0.333em \xspaceskip=0.5em }
\def\titlestyle#1{\par\begingroup \titleparagraphs
     \iftwelv@\fourteenpoint\else\twelvepoint\fi
   \noindent #1\par\endgroup }
\def\spacecheck#1{\dimen@=\pagegoal\advance\dimen@ by -\pagetotal
   \ifdim\dimen@<#1 \ifdim\dimen@>0pt \vfil\break \fi\fi}
\def\chapter#1{\par \penalty-300 \vskip\chapterskip
   \spacecheck\chapterminspace
   \chapterreset \titlestyle{\chapterlabel.~#1}
   \nobreak\vskip\headskip \penalty 30000
   {\pr@tect\wlog{\string\chapter\space \chapterlabel}} }

\def\section#1{\par \ifnum\the\lastpenalty=30000\else
   \penalty-200\vskip\sectionskip \spacecheck\sectionminspace\fi
   \gl@bal\advance\sectionnumber by 1
   {\pr@tect
   \xdef\sectionlabel{\the\sectionstyle\the\sectionnumber}
   \wlog{\string\section\space \sectionlabel}}
   \noindent {\caps\enspace\sectionlabel.~~#1}\par
   \nobreak\vskip\headskip \penalty 30000 }
\def\subsection#1{\par
   \ifnum\the\lastpenalty=30000\else \penalty-100\smallskip \fi
   \noindent\undertext{#1}\enspace \vadjust{\penalty5000}}

\def\undertext#1{\vtop{\hbox{#1}\kern 1pt \hrule}}
\def\APPENDIX#1#2{\par\penalty-300\vskip\chapterskip
   \spacecheck\chapterminspace \chapterreset \xdef\chapterlabel{#1}
   \titlestyle{APPENDIX #2} \nobreak\vskip\headskip \penalty 30000
   \wlog{\string\Appendix~\chapterlabel} }
\def\Appendix#1{\APPENDIX{#1}{#1}}
\def\appendix{\APPENDIX{A}{}}
\def\unnumberedchapters{\let\makechapterlabel=\rel@x
      \let\chapterlabel=\rel@x  \sectionstyle={\BLANC}
      \let\sectionlabel=\rel@x \sequentialequations }
%
%
%
\def\eqname#1{\rel@x {\pr@tect
  \ifnum\equanumber<0 \xdef#1{{\rm(\number-\equanumber)}}%
     \gl@bal\advance\equanumber by -1
  \else \gl@bal\advance\equanumber by 1
     \ifx\chapterlabel\rel@x \def\d@t{}\else \def\d@t{.}\fi
    \xdef#1{{\rm(\chapterlabel\d@t\number\equanumber)}}\fi #1}}

\def\eqn{\eqno\eqname}

\def\eqinsert#1{\noalign{\dimen@=\prevdepth \nointerlineskip
   \setbox0=\hbox to\displaywidth{\hfil #1}
   \vbox to 0pt{\kern 0.5\baselineskip\hbox{$\!\box0\!$}\vss}
   \prevdepth=\dimen@}}
%

%
%
\def\GENITEM#1;#2{\par \hangafter=0 \hangindent=#1
    \Textindent{$ #2 $}\ignorespaces}
\outer\def\newitem#1=#2;{\gdef#1{\GENITEM #2;}}

\newdimen\itemsize                \itemsize=30pt
\newitem\item=1\itemsize;
\newitem\sitem=1.75\itemsize;     
\newitem\ssitem=2.5\itemsize;     
\outer\def\newlist#1=#2&#3&#4;{\toks0={#2}\toks1={#3}%
   \count255=\escapechar \escapechar=-1
   \alloc@0\list\countdef\insc@unt\listcount     \listcount=0
   \edef#1{\par
      \countdef\listcount=\the\allocationnumber
      \advance\listcount by 1
      \hangafter=0 \hangindent=#4
      \Textindent{\the\toks0{\listcount}\the\toks1}}
   \expandafter\expandafter\expandafter
    \edef\c@t#1{begin}{\par
      \countdef\listcount=\the\allocationnumber \listcount=1
      \hangafter=0 \hangindent=#4
      \Textindent{\the\toks0{\listcount}\the\toks1}}
   \expandafter\expandafter\expandafter
    \edef\c@t#1{con}{\par \hangafter=0 \hangindent=#4 \noindent}
   \escapechar=\count255}
\def\c@t#1#2{\csname\string#1#2\endcsname}
\newlist\point=\Number&.&1.0\itemsize;
\newlist\subpoint=(\alphabetic&)&1.75\itemsize;
\newlist\subsubpoint=(\roman&)&2.5\itemsize;
%

%
%
%
%
\newcount\referencecount     \referencecount=0
\newcount\lastrefsbegincount \lastrefsbegincount=0
\newif\ifreferenceopen       \newwrite\referencewrite
\newdimen\refindent          \refindent=30pt
\def\normalrefmark#1{\attach{\scriptscriptstyle [ #1 ] }}
\let\PRrefmark=\attach
\def\NPrefmark#1{\step@ver{{\;[#1]}}}
\def\refmark#1{\rel@x\ifPhysRev\PRrefmark{#1}\else\normalrefmark{#1}\fi}
\def\refend@{\refmark{\number\referencecount}}
\def\refend{\refend@{}\space }
\def\refsend{\refmark{\count255=\referencecount
   \advance\count255 by-\lastrefsbegincount
   \ifcase\count255 \number\referencecount
   \or \number\lastrefsbegincount,\number\referencecount
   \else \number\lastrefsbegincount-\number\referencecount \fi}\space }
\def\REFNUM#1{\rel@x \gl@bal\advance\referencecount by 1
    \xdef#1{\the\referencecount }}
\def\Refnum#1{\REFNUM #1\refend@ } 
\def\REF#1{\REFNUM #1\R@FWRITE\ignorespaces}
\def\Ref#1{\Refnum #1\REFWRITE }
\def\ref{\Ref\?}
\def\REFS#1{\REFNUM #1\gl@bal\lastrefsbegincount=\referencecount
    \REFWRITE }

\def\r@fitem#1{\par \hangafter=0 \hangindent=\refindent \Textindent{#1}}
\def\refitem#1{\r@fitem{#1.}}
\def\NPrefitem#1{\r@fitem{[#1]}}
\def\NPrefs{\let\refmark=\NPrefmark \let\refitem=\NPrefitem}
\def\REFWRITE{\R@FWRITE\rel@x }
\def\R@FWRITE#1{\ifreferenceopen \else \gl@bal\referenceopentrue
     \immediate\openout\referencewrite=\jobname.refs
     \toks@={\begingroup \refoutspecials \catcode`\^^M=10 }%
     \immediate\write\referencewrite{\the\toks@}\fi
    \immediate\write\referencewrite{\noexpand\refitem %
                                    {\the\referencecount}}%
    \p@rse@ndwrite \referencewrite #1}
\begingroup
 \catcode`\^^M=\active \let^^M=\relax %
 \gdef\p@rse@ndwrite#1#2{\begingroup \catcode`\^^M=12 \newlinechar=`\^^M%
         \chardef\rw@write=#1\sc@nlines#2}%
 \gdef\sc@nlines#1#2{\sc@n@line \g@rbage #2^^M\endsc@n \endgroup #1}%
 \gdef\sc@n@line#1^^M{\expandafter\toks@\expandafter{\deg@rbage #1}%
         \immediate\write\rw@write{\the\toks@}%
         \futurelet\n@xt \sc@ntest }%
\endgroup
\def\sc@ntest{\ifx\n@xt\endsc@n \let\n@xt=\rel@x
       \else \let\n@xt=\sc@n@notherline \fi \n@xt }
\def\sc@n@notherline{\sc@n@line \g@rbage }
\def\deg@rbage#1{}
\let\g@rbage=\relax    \let\endsc@n=\relax
\def\refout{\par\penalty-400\vskip\chapterskip
   \spacecheck\referenceminspace
   \ifreferenceopen \Closeout\referencewrite \referenceopenfalse \fi
   \line{\fourteenrm\hfil REFERENCES\hfil}\vskip\headskip
   \input \jobname.refs
   }
\def\refoutspecials{\sfcode`\.=1000 \interlinepenalty=1000
         \rightskip=\z@ plus 1em minus \z@ }
\def\Closeout#1{\toks0={\par\endgroup}\immediate\write#1{\the\toks0}%
   \immediate\closeout#1}
%
%
\newcount\figurecount     \figurecount=0
\newcount\tablecount      \tablecount=0
\newif\iffigureopen       \newwrite\figurewrite
\newif\iftableopen        \newwrite\tablewrite
\def\FIGNUM#1{\rel@x \gl@bal\advance\figurecount by 1
    \xdef#1{\the\figurecount}}
\def\FIGURE#1{\FIGNUM #1\F@GWRITE\ignorespaces }

\def\figitem#1{\r@fitem{#1)}}
\def\FIGWRITE{\F@GWRITE\rel@x }
\def\TABNUM#1{\rel@x \gl@bal\advance\tablecount by 1
    \xdef#1{\the\tablecount}}
\def\TABLE#1{\TABNUM #1\T@BWRITE\ignorespaces }

\def\tabitem#1{\r@fitem{#1:}}
\def\TABWRITE{\T@BWRITE\rel@x }
\def\F@GWRITE#1{\iffigureopen \else \gl@bal\figureopentrue
     \immediate\openout\figurewrite=\jobname.figs
     \toks@={\begingroup \catcode`\^^M=10 }%
     \immediate\write\figurewrite{\the\toks@}\fi
    \immediate\write\figurewrite{\noexpand\figitem %
                                 {\the\figurecount}}%
    \p@rse@ndwrite \figurewrite #1}
\def\T@BWRITE#1{\iftableopen \else \gl@bal\tableopentrue
     \immediate\openout\tablewrite=\jobname.tabs
     \toks@={\begingroup \catcode`\^^M=10 }%
     \immediate\write\tablewrite{\the\toks@}\fi
    \immediate\write\tablewrite{\noexpand\tabitem %
                                 {\the\tablecount}}%
    \p@rse@ndwrite \tablewrite #1}
\def\figout{\par\penalty-400
   \vskip\chapterskip\spacecheck\referenceminspace
   \iffigureopen \Closeout\figurewrite \figureopenfalse \fi
   \line{\fourteenrm\hfil FIGURE CAPTIONS\hfil}\vskip\headskip
   \input \jobname.figs
   }
\def\tabout{\par\penalty-400
   \vskip\chapterskip\spacecheck\referenceminspace
   \iftableopen \Closeout\tablewrite \tableopenfalse \fi
   \line{\fourteenrm\hfil TABLE CAPTIONS\hfil}\vskip\headskip
   \input \jobname.tabs
   }
%
%
%
\newbox\picturebox
\def\p@cht{\ht\picturebox }
\def\p@cwd{\wd\picturebox }
\def\p@cdp{\dp\picturebox }
\newdimen\xshift
\newdimen\yshift
\newdimen\captionwidth
\newskip\captionskip
\captionskip=15pt plus 5pt minus 3pt
\def\fullwidth{\captionwidth=\hsize }
\newtoks\Caption
\newif\ifcaptioned
\newif\ifselfcaptioned
\def\caption{\captionedtrue \Caption }
\newcount\linesabove
\newif\iffileexists
\newtoks\picfilename
\def\fil@#1 {\fileexiststrue \picfilename={#1}}
\def\file#1{\if=#1\let\n@xt=\fil@ \else \def\n@xt{\fil@ #1}\fi \n@xt }
\def\pl@t{\begingroup \pr@tect
    \setbox\picturebox=\hbox{}\fileexistsfalse
    \let\height=\p@cht \let\width=\p@cwd \let\depth=\p@cdp
    \xshift=\z@ \yshift=\z@ \captionwidth=\z@
    \Caption={}\captionedfalse
    \linesabove =0 \picturedefault }
\def\plot{\pl@t \selfcaptionedfalse }
\def\Picture#1{\gl@bal\advance\figurecount by 1
    \xdef#1{\the\figurecount}\pl@t \selfcaptionedtrue }

\def\s@vepicture{\iffileexists \parsefilename \redopicturebox \fi
   \ifdim\captionwidth>\z@ \else \captionwidth=\p@cwd \fi
   \xdef\lastpicture{\iffileexists
        \setbox0=\hbox{\raise\the\yshift \vbox{%
              \moveright\the\xshift\hbox{\picturedefinition}}}%
        \else \setbox0=\hbox{}\fi
         \ht0=\the\p@cht \wd0=\the\p@cwd \dp0=\the\p@cdp
         \vbox{\hsize=\the\captionwidth \line{\hss\box0 \hss }%
              \ifcaptioned \vskip\the\captionskip \noexpand\Tenpoint
                \ifselfcaptioned Figure~\the\figurecount.\enspace \fi
                \the\Caption \fi }}%
    \endgroup }
\let\endpicture=\s@vepicture
\def\savepicture#1{\s@vepicture \global\let#1=\lastpicture }
\def\displaypicture{\fullwidth \s@vepicture $$\lastpicture $${}}
\def\toppicture{\fullwidth \s@vepicture \topinsert
    \lastpicture \medskip \endinsert }
\def\midpicture{\fullwidth \s@vepicture \midinsert
    \lastpicture \endinsert }
%
%
\def\leftpicture{\pres@tpicture
    \dimen@i=\hsize \advance\dimen@i by -\dimen@ii
    \setbox\picturebox=\hbox to \hsize {\box0 \hss }%
    \wr@paround }
\def\rightpicture{\pres@tpicture
    \dimen@i=\z@
    \setbox\picturebox=\hbox to \hsize {\hss \box0 }%
    \wr@paround }
\def\pres@tpicture{\gl@bal\linesabove=\linesabove
    \s@vepicture \setbox\picturebox=\vbox{
         \kern \linesabove\baselineskip \kern 0.3\baselineskip
         \lastpicture \kern 0.3\baselineskip }%
    \dimen@=\p@cht \dimen@i=\dimen@
    \advance\dimen@i by \pagetotal
    \par \ifdim\dimen@i>\pagegoal \vfil\break \fi
    \dimen@ii=\hsize
    \advance\dimen@ii by -\parindent \advance\dimen@ii by -\p@cwd
    \setbox0=\vbox to\z@{\kern-\baselineskip \unvbox\picturebox \vss }}
\def\wr@paround{\Caption={}\count255=1
    \loop \ifnum \linesabove >0
         \advance\linesabove by -1 \advance\count255 by 1
         \advance\dimen@ by -\baselineskip
         \expandafter\Caption \expandafter{\the\Caption \z@ \hsize }%
      \repeat
    \loop \ifdim \dimen@ >\z@
         \advance\count255 by 1 \advance\dimen@ by -\baselineskip
         \expandafter\Caption \expandafter{%
             \the\Caption \dimen@i \dimen@ii }%
      \repeat
    \edef\n@xt{\parshape=\the\count255 \the\Caption \z@ \hsize }%
    \par\noindent \n@xt \strut \vadjust{\box\picturebox }}
\let\picturedefault=\relax
\let\parsefilename=\relax
\def\redopicturebox{\let\picturedefinition=\rel@x
   \errhelp=\disabledpictures
   \errmessage{This version of TeX cannot handle pictures.  Sorry.}}
\newhelp\disabledpictures
     {You will get a blank box in place of your picture.}
%
%
%
%
%
%
%
%
%
%
\def\FRONTPAGE{\ifvoid255\else\vfill\penalty-20000\fi
   \gl@bal\pagenumber=1     \gl@bal\chapternumber=0
   \gl@bal\equanumber=0     \gl@bal\sectionnumber=0
   \gl@bal\referencecount=0 \gl@bal\figurecount=0
   \gl@bal\tablecount=0     \gl@bal\frontpagetrue
   \gl@bal\lastf@@t=0       \gl@bal\footsymbolcount=0}

\def\papers{\papersize\headline=\paperheadline\footline=\paperfootline}
\def\papersize{
   \advance\hoffset by\HOFFSET \advance\voffset by\VOFFSET
   \pagebottomfiller=0pc
   \skip\footins=\bigskipamount \normalspace }
\papers  
%
%
\newskip\lettertopskip       \lettertopskip=20pt plus 50pt
\newskip\letterbottomskip    \letterbottomskip=\z@ plus 100pt
\newskip\signatureskip       \signatureskip=40pt plus 3pt
\def\lettersize{\hsize=6.5in \vsize=8.5in \hoffset=0in \voffset=0.5in
   \advance\hoffset by\HOFFSET \advance\voffset by\VOFFSET
   \pagebottomfiller=\letterbottomskip
   \skip\footins=\smallskipamount \multiply\skip\footins by 3
   \singlespace }
\def\MEMO{\lettersize \headline=\letterheadline \footline={\hfil }%
   \let\rule=\memorule \FRONTPAGE \memohead }

\def\memodate{\afterassignment\MEMO \date }
\def\memit@m#1{\smallskip \hangafter=0 \hangindent=1in
    \Textindent{\caps #1}}
\def\subject{\memit@m{Subject:}}
\def\topic{\memit@m{Topic:}}
\def\from{\memit@m{From:}}
\def\memorule{\medskip\hrule height 1pt\bigskip}  
\def\memohead{\centerline{\fourteenrm MEMORANDUM}}
\newwrite\labelswrite
\newtoks\rw@toks
\def\letters{\lettersize
   \headline=\letterheadline \footline=\letterfootline
   \immediate\openout\labelswrite=\jobname.lab}

\let\letterhead=\rel@x
\def\addressee#1{\medskip\line{\hskip 0.75\hsize plus\z@ minus 0.25\hsize
                               \the\date \hfil }%
   \vskip \lettertopskip
   \ialign to\hsize{\strut ##\hfil\tabskip 0pt plus \hsize \crcr #1\crcr}
   \writelabel{#1}\medskip \noindent\hskip -\spaceskip \ignorespaces }
\def\rwl@begin#1\cr{\rw@toks={#1\crcr}\rel@x
   \immediate\write\labelswrite{\the\rw@toks}\futurelet\n@xt\rwl@next}
\def\rwl@next{\ifx\n@xt\rwl@end \let\n@xt=\rel@x
      \else \let\n@xt=\rwl@begin \fi \n@xt}
\let\rwl@end=\rel@x
\def\writelabel#1{\immediate\write\labelswrite{\noexpand\labelbegin}
     \rwl@begin #1\cr\rwl@end
     \immediate\write\labelswrite{\noexpand\labelend}}
\newtoks\FromAddress         \FromAddress={}
\newtoks\sendername          \sendername={}
\newbox\FromLabelBox
\newdimen\labelwidth          \labelwidth=6in
\def\makelabels{\afterassignment\Makelabels \sendersname=}
\def\Makelabels{\FRONTPAGE \letterinfo={\hfil } \MakeFromBox
     \immediate\closeout\labelswrite  \input \jobname.lab\vfil\eject}
\let\labelend=\rel@x
\def\labelbegin#1\labelend{\setbox0=\vbox{\ialign{##\hfil\cr #1\crcr}}
     \MakeALabel }
\def\MakeFromBox{\gl@bal\setbox\FromLabelBox=\vbox{\Tenpoint
     \ialign{##\hfil\cr \the\sendername \the\FromAddress \crcr }}}
\def\MakeALabel{\vskip 1pt \hbox{\vrule \vbox{
        \hsize=\labelwidth \hrule\bigskip
        \leftline{\hskip 1\parindent \copy\FromLabelBox}\bigskip
        \centerline{\hfil \box0 } \bigskip \hrule
        }\vrule } \vskip 1pt plus 1fil }
\def\signed#1{\par \nobreak \bigskip \dt@pfalse \begingroup
  \everycr={\noalign{\nobreak
            \ifdt@p\vskip\signatureskip\gl@bal\dt@pfalse\fi }}%
  \tabskip=0.5\hsize plus \z@ minus 0.5\hsize
  \halign to\hsize {\strut ##\hfil\tabskip=\z@ plus 1fil minus \z@\crcr
          \noalign{\gl@bal\dt@ptrue}#1\crcr }%
  \endgroup \bigskip }
\newbox\letterb@x
\def\lettertext{\par \vskip\parskip \unvcopy\letterb@x \par }
\def\multiletter{\setbox\letterb@x=\vbox\bgroup
      \everypar{\vrule height 1\baselineskip depth 0pt width 0pt }
      \singlespace \topskip=\baselineskip }
\def\letterend{\par\egroup}
%
%
%
\newskip\frontpageskip
\newtoks\Pubnum   
\newtoks\Pubtype  \let\pubtype=\Pubtype
\newif\ifp@bblock  \p@bblocktrue
\def\PH@SR@V{\doubl@true \baselineskip=24.1pt plus 0.2pt minus 0.1pt
             \parskip= 3pt plus 2pt minus 1pt }
\def\PHYSREV{\papers\PhysRevtrue\PH@SR@V}

\def\titlepage{\FRONTPAGE\papers\ifPhysRev\PH@SR@V\fi
   \ifp@bblock\p@bblock \else\hrule height\z@ \rel@x \fi }
\def\nopubblock{\p@bblockfalse}

\frontpageskip=12pt plus .5fil minus 2pt
\Pubtype={}
\Pubnum={}
\def\p@bblock{\begingroup \tabskip=\hsize minus \hsize
   \baselineskip=1.5\ht\strutbox \topspace-2\baselineskip
   \halign to\hsize{\strut ##\hfil\tabskip=0pt\crcr
       \the\Pubnum\crcr\the\date\crcr\the\pubtype\crcr}\endgroup}
\def\title#1{\vskip\frontpageskip \titlestyle{#1} \vskip\headskip }
\def\author#1{\vskip\frontpageskip\titlestyle{\twelvecp #1}\nobreak}

\def\address#1{\par\kern 5pt\titlestyle{\twelvepoint\it #1}}
\def\andaddress{\par\kern 5pt \centerline{\sl and} \address}

\def\abstract{\par\dimen@=\prevdepth \hrule height\z@ \prevdepth=\dimen@
   \vskip\frontpageskip\centerline{\fourteenrm ABSTRACT}\vskip\headskip }

%
%
%

\def\etal{\hbox{\it et al.}}   
\def\\{\rel@x \ifmmode \backslash \else {\tt\char`\\}\fi }
\def\sequentialequations{\rel@x \if\equanumber<0 \else
  \gl@bal\equanumber=-\equanumber \gl@bal\advance\equanumber by -1 \fi }
\def\journal#1&#2(#3){\begingroup \let\journal=\dummyj@urnal
    \unskip, \sl #1\unskip~\bf\ignorespaces #2\rm
    (\afterassignment\j@ur \count255=#3), \endgroup\ignorespaces }
\def\j@ur{\ifnum\count255<100 \advance\count255 by 1900 \fi
          \number\count255 }
\def\dummyj@urnal{%
    \toks@={Reference foul up: nested \journal macros}%
    \errhelp={Your forgot & or ( ) after the last \journal}%
    \errmessage{\the\toks@ }}

\def\topspace{\hrule height 0pt depth 0pt \vskip}

\def\Buildrel#1\under#2{\mathrel{\mathop{#2}\limits_{#1}}}
\def\becomes#1{\mathchoice{\becomes@\scriptstyle{#1}}
   {\becomes@\scriptstyle{#1}} {\becomes@\scriptscriptstyle{#1}}
   {\becomes@\scriptscriptstyle{#1}}}
\def\becomes@#1#2{\mathrel{\setbox0=\hbox{$\m@th #1{\,#2\,}$}%
        \mathop{\hbox to \wd0 {\rightarrowfill}}\limits_{#2}}}

\let\int=\intop         
\def\lsim{\mathrel{\mathpalette\@versim<}}
\def\gsim{\mathrel{\mathpalette\@versim>}}
\def\@versim#1#2{\vcenter{\offinterlineskip
        \ialign{$\m@th#1\hfil##\hfil$\crcr#2\crcr\sim\crcr } }}
\def\big#1{{\hbox{$\left#1\vbox to 0.85\b@gheight{}\right.\n@space$}}}
\def\Big#1{{\hbox{$\left#1\vbox to 1.15\b@gheight{}\right.\n@space$}}}
\def\bigg#1{{\hbox{$\left#1\vbox to 1.45\b@gheight{}\right.\n@space$}}}
\def\Bigg#1{{\hbox{$\left#1\vbox to 1.75\b@gheight{}\right.\n@space$}}}
\def\){\mskip 2mu\nobreak }
%
%
%
\let\sec@nt=\sec
\def\sec{\rel@x\ifmmode\let\n@xt=\sec@nt\else\let\n@xt\section\fi\n@xt}
\def\obsolete#1{\message{Macro \string #1 is obsolete.}}
\def\firstsec#1{\obsolete\firstsec \section{#1}}
\def\firstsubsec#1{\obsolete\firstsubsec \subsection{#1}}
\def\thispage#1{\obsolete\thispage \gl@bal\pagenumber=#1\frontpagefalse}
\def\thischapter#1{\obsolete\thischapter \gl@bal\chapternumber=#1}
\def\splitout{\obsolete\splitout\rel@x}
\def\prop{\obsolete\prop \propto }
\def\nextequation#1{\obsolete\nextequation \gl@bal\equanumber=#1
   \ifnum\the\equanumber>0 \gl@bal\advance\equanumber by 1 \fi}
\def\BOXITEM{\afterassigment\B@XITEM\setbox0=}
\def\B@XITEM{\par\hangindent\wd0 \noindent\box0 }
%
%
%
\def\phyzzx{PHY\setbox0=\hbox{Z}\copy0 \kern-0.5\wd0 \box0 X}
        
\everyjob{\xdef\today{\monthname~\number\day, \number\year}
        \input myphyx.tex }
\message{ by V.K.}
%
\catcode`\@=12 
%

%
%
%
\baselineskip=18.5pt plus 1pt\lineskip=2pt\lineskiplimit=1pt
\def\etal{et al. }
\centerline{\bf AN ANALYTICAL MODEL FOR THE TRIAXIAL COLLAPSE}
\centerline{\bf OF COSMOLOGICAL PERTURBATIONS}
\bigskip
\centerline{Daniel J. Eisenstein\footnote{1}{Also at:
Physics Department, Harvard University} and Abraham Loeb}
\centerline{Astronomy Department, Harvard University}
\centerline{60 Garden St., Cambridge MA 02138}
\bigskip

\centerline{\bf ABSTRACT}

We present an analytical model for the non-spherical collapse of
overdense regions out of a Gaussian random field of initial cosmological 
perturbations.  
The collapsing region is treated
as an ellipsoid of constant density, acted upon by the quadrupole tidal shear 
from the surrounding matter.  The dynamics of the ellipsoid is
set by the ellipsoid self-gravity and the external quadrupole shear.
Both forces are linear in the
coordinates and therefore maintain homogeneity of the 
ellipsoid at all times.
The amplitude of the external shear is evolved into
the non-linear regime in thin spherical shells that are allowed to 
move only radially according to the mass interior to them.
The full dynamical equations then reduce
to a set of nine second-order ordinary differential equations,
which reproduce the linear regime behaviour 
but can be evolved past turnaround, well into the non-linear regime.
We describe how the initial conditions can be drawn in the appropriate
correlated way from a random field
of initial density perturbations.
The model is applied to a restricted set of initial conditions 
that are more suitable to the above approximations; most
notably we focus on the properties of rare high density peaks ($\gsim 2\sigma$).
By considering many random realizations of the initial conditions,
we calculate the distribution of shapes and angular momenta
acquired by objects through the coupling of their
quadrupole moment to the tidal shear. 
The average value of 
the spin parameter, $\langle\lambda\rangle\approx 0.04$, is found to be 
only weakly dependent on the system mass, the mean 
cosmological density, or the initial power 
spectrum of perturbations,
in agreement with N-body simulations. 
For the cold dark matter power spectrum, most objects evolve 
from a quasi-spherical initial state
to a pancake or filament and then to complete virialization.
Low-spin objects tend to be more spherical.
The evolution history of shapes is primarily induced by the external shear and not 
by the initial triaxiality of the objects. 
The statistical distribution of the triaxial shapes 
of collapsing regions can be used to test cosmological models
against galaxy surveys on large scales. 

\noindent
{\it Subject headings:} cosmology: theory

\np
\centerline{\bf 1. INTRODUCTION}

In the standard cosmological model, density perturbations grow
by gravitational instability from their small initial amplitude to
the observed non-linear structure, thus accounting for the galaxies,
clusters, superclusters, filaments and voids in the present universe (Peebles 1993).
Despite the simplicity of these initial conditions, the analytical 
understanding of the 
advanced stages of gravitational instability,
where the density contrast $\delta\rho/\rho$ exceeds unity,
is limited.
Numerical codes are frequently 
used in an attempt to uncover the non-linear dynamics of collapsing regions
and in order 
to predict the observational consequences of specific models for 
structure formation.
However,  simulations are often expensive in computer time 
and are therefore limited in resolution or total volume.
An approximate extension of 
linear perturbation theory into the non-linear regime 
is achieved by the Zel'dovich (1970) approximation (see review by Shandarin \& Zel'dovich
1989). 
Another approach which extrapolates the physics even further,
up to the virialization phase of bound objects, is the spherical collapse model 
(Gunn \& Gott, 1972; Peebles 1980, \S19).  
Under the assumption of sphericity, the non-linear dynamics of 
a collapsing shell is 
determined by the mass interior to it 
and described by the parametric solution to
the dynamics of a closed universe.
Although the Zel'dovich approximation generically yields 
pancakes and  
numerical simulations show that 
many collapsing regions are filamentary (e.g. Park 1990; Bertschinger \& Gelb 1991; 
Cen \& Ostriker 1993),
the spherical model became popular because of its simplicity.
Most notably, it was integrated
into the Press-Schechter formalism (Press \& Schechter 1974) for calculating
the mass function of collapsed objects in the universe. In reality,
overdense regions in galaxy surveys of
the local universe (e.g., Geller \& Huchra 1989; Maddox et al. 1990; Saunders
et al. 1991; Shectman et al. 1992; Strauss et al. 1992)
have complicated triaxial shapes with sheets 
and filaments being common features.  Moreover, recent
work has demonstrated that the tidal shear 
plays an important role in the dynamics of collapsing regions 
(Dubinski 1992; Bond \& Myers 1993; Bertschinger \& Jain 1994; 
van de Weygaert \& Babul 1994) 
and therefore requires a non-spherical analysis.  
Non-sphericity is also necessary in order to explain the origin of galactic rotation 
through the coupling of galaxies to tidal torques during their collapse 
(Hoyle 1949; Peebles 1969; Doroshkevich 1970;
Efstathiou \& Jones 1979; White 1984; Hoffman 1986;
Barnes \& Efstathiou 1987; Ryden 1988; Quinn \& Binney 1992; 
Warren \etal 1992). 

In this paper we construct a new analytical 
model to study the non-linear collapse of non-spherical regions
in a Gaussian random field of initial density perturbations.  
The collapsing system is approximated as a triaxial ellipsoid of 
constant overdensity that is affected by its own gravitational field and by the 
external tidal shear. The external 
tidal force can be expanded as a multipole series,
with the quadrupole being the dominant relevant term
for the internal dynamics of the ellipsoid. 
Since the quadrupole force and the ellipsoid self-gravity are linear 
functions of the spatial coordinates, they maintain    
homogeneity
of the ellipsoid at all times.
The quadrupole shear from the external density field
is calculated by dividing the background
mass distribution into spherical shells that move only radially.
Under these approximations, we reduce the full dynamical equations of
motion to a set of second-order ordinary differential equations.
The initial conditions can be derived from a Gaussian random field of density 
perturbations with some power spectrum, and the initial peculiar 
velocities are chosen to be those of the
growing mode. The equations of motion we obtain reproduce 
the linear regime behaviour of the density field.
By integrating these differential equations we are able to describe
the collapse of a triaxial sheared region into
a virialized object. In difference from the spherical
approach, the ellipsoid model can
be used to study the influence of the external 
tidal shear on the triaxial collapse of overdense regions. 
Through many random realizations of the initial density field
the model can 
examine the statistical properties of collapsing systems,
including their triaxial shapes, orientation relative to the background
density field, and total angular momenta,  
for different cosmological models.

The outline of this work is as follows. In \S 2 we review the 
equations of motion for a self-gravitating ellipsoid
and the approximation (Icke 1973)
that allows us to extend this
treatment to a homogeneous ellipsoidal overdensity in the universe.  
We also show that 
the quadrupole shear can be added to this formulation. In \S 3 we present our
approximation for the non-linear evolution of this shear.    
The restrictions placed upon the initial conditions for the
model are described in \S 4. Given a Gaussian random field
of initial density perturbations, we derive in Appendix A
the joint 
probability distribution 
for the necessary initial conditions.
In Appendix B we describe our treatment of a collapse along 
a single axis, and in 
Appendix C we relate the magnitude of the external shear 
to the amplitude of density fluctuations on a given mass scale in a fashion independent of
the initial power spectrum.
The statistical properties of collapsing regions are then
analyzed 
in \S 5, by applying 
the above model to many realizations of random initial conditions.
Finally, in \S 6 we discuss the applications and limitations of the model
and indicate potential future work.

\np
\centerline{\bf 2. ELLIPSOID MODEL FOR A COLLAPSING REGION}

We begin from a small density perturbation on top of a smooth background
assumed to be a Friedmann-Robertson-Walker (FRW) universe.  
We assume that the background is composed of collisionless cold particles
and allow for a non-zero cosmological constant.
We pick the origin to be
near the center of a region that has a sufficiently high density so that 
it collapses before its environment.  Since we are interested 
in the properties of the collapsing region, we separate the universe into
two disjoint parts, the collapsing high density region and the rest of the
universe.  
The boundary between these parts is taken to be a sphere centered at the origin.
We denote the background density by $\rho_b$ and 
let the density as a function of space
be $\rho({\bf x})$, where $\bf x$ is the position vector. We then define 
$\delta\rho({\bf x}) = \rho({\bf x}) - \rho_b$
and $\delta({\bf x}) = \delta\rho({\bf x})/\rho_b$.

The gravitational force exerted on the spherical region by the rest of the universe
can be calculated 
by expanding the external gravitational potential 
as a multipole series (Binney and Tremaine 1987).  
Taking the sphere to have radius $R$, we find
$$
\Phi({\bf x}) = \sum_{\ell,m}{{4\pi G\over 2\ell+1} a_{\ell m} Y_{\ell m} x^\ell} ,
\eqn\labe
$$
where $x \equiv |{\bf x}|$, the $Y_{\ell m}$ are spherical harmonics, $G$ is Newton's
constant, and 
$$
a_{\ell m} = \rho_b \int_{|{\bf s}|>R}{d^3s\, Y^*_{\ell m} \delta({\bf s}) s^{-\ell-1}}.
\eqn\labf
$$
The time-dependent magnitude of the external potential
will be calculated in \S 3.
For now we consider the effect of this
potential on the inner region.

We are primarily concerned with the acquisition of angular momentum and the 
shearing of the inner region.  For these purposes, we focus on the 
quadrupole ($\ell = 2$) terms.  
The $\ell =0$ term produces no force, and the dipole
($\ell = 1$) terms produce a uniform acceleration that may move the whole inner region
but does not alter the shape or induce any rotation.
This motion can indirectly affect the object because the surrounding material will
have a new angular distribution relative to the displaced object, requiring a re-calculation
of the multipole expansion coefficients.  However, if the dipole is generated on large
scales, then the object and its entire 
neighborhood move together as a bulk flow and the changes in the 
angular distribution of matter will be very small, 
allowing us to ignore the $\ell=1$ terms. 
Previous work has found the quadrupole terms to dominate  
the higher ($\ell\ge 3$) terms (Quinn \& Binney 1992), and
so we ignore all but the $\ell = 2$ terms.

We model the inner region as a homogeneous ellipsoidal overdensity
(Lynden-Bell 1964; Lin, Mestel \& Shu 1965; Zel'dovich 1965; Icke 1973; White \& Silk 1979). 
By this, we mean that $\delta({\bf x})$ is a constant inside the ellipsoid 
and zero outside of it.  Homogeneous ellipsoids are advantageous because their
gravitational potentials depend quadratically on the coordinates and therefore 
preserve their homogeneity 
and because they have non-zero quadrupole moments which
couple to the $\ell = 2$ terms of the external potential.  Obviously, this
approximation ignores all the complexities of shapes and substructure actually
present in the central region.  However the object
cannot be torqued by its sub-components,
and the time dependence of the 
quadrupole moments may
be well captured by the ellipsoid approximation.

A convenient way to analyze the motion of homogeneous ellipsoids was discussed
by Peebles (1980, \S20).  The position of the ellipsoid is given by 
$$
r^\alpha = A^{\alpha\beta}x^\beta,
\eqn\labb
$$
where ${\bf x}$ is a vector inside the unit sphere and repeated indices are summed.  
The matrix 
$A$ is constant in space but depends on time.  
The mass is evenly distributed
over this unit sphere and thus the ellipsoid has a uniform density.  If the columns of
$A$ are orthogonal, then they are the axes of the ellipsoid, with the 
lengths of the axes being the magnitude of the respective column.  
However, since in general
the columns of $A$ are not orthogonal, the axes may be found
from the relation defining the outer shell of the ellipsoid,  
$x^\alpha x^\alpha = 1$.  The equation of the ellipsoid is therefore 
$$
{\bf r}^T A^{-1T} A^{-1} {\bf r} = 1 .
\eqn\laba
$$
To rotate to the principal axis frame
we diagonalize the matrix $AA^T$ as $Q\Lambda Q^T$,
where $Q$ is orthogonal and $\Lambda$ is real and diagonal
because $AA^T$ is symmetric.
Then we rewrite equation $\laba$ as 
$$(Q^T{\bf r})^T\Lambda^{-1}(Q^T{\bf r}) = 1.\eqn\labza$$
Thus, the axes of the ellipsoid are in the columns of Q and the corresponding
axis length is the square root of the corresponding diagonal element of $\Lambda$.

Since the gravitational potential of such an
ellipsoid may be written as a quadratic function (Peebles 1980),
we are prompted to consider a general quadratic potential
$$
\Phi({\bf r}) = {1\over2} \Phi^{\alpha\beta}r^{\alpha}r^{\beta},\eqn\labzb
$$
where the matrix $\Phi^{\alpha\beta}$ is a function of the matrix $A$.  
Then the force per unit mass is just $-\nabla\Phi({\bf r})$, leading
to the momentum equation,
$$
{d^2r^\alpha\over dt^2} = -\Phi^{\alpha\beta}r^\beta .
\eqn\mot
$$
By substituting equation $\labb$ in $\mot$ we find
$$
{d^2A^{\alpha\beta}\over dt^2} = -\Phi^{\alpha\gamma}A^{\gamma\beta}.
\eqn\labg
$$
This is our basic equation of motion.
It is crucial that the potential be quadratic in the coordinates, since
this leads to forces that
are linear in space.  The magnitude of ${\bf r}$ cancels out, so that all the similar
ellipsoidal shells behave in the same way and the ellipsoid remains homogeneous. 

We now must construct this quadratic potential.
First, we consider the contribution from the ellipsoid.
For an isolated ellipsoid of mass $M_e$, the potential is given
by elliptic functions (Peebles 1980) so that,
$$
\Phi_{ell}({\bf r'}) = {1\over2}G M_e \left[(r'_1)^2 R_D(a_2^2,a_3^2,a_1^2) +
(r'_2)^2 R_D(a_1^2,a_3^2,a_2^2) + (r'_3)^2 R_D(a_1^2,a_2^2,a_3^2)\right],\eqn\labzc
$$
where $a_j$ are the lengths of the semi-axes and ${\bf r'}$ is in the principal
axis frame.  The function $R_D$ is defined as (Carlson 1977; Press \etal 1992), 
$$
R_D(x,y,z) = {3\over2} \int^\infty_0{dt\over(t+z)\sqrt{(t+x)(t+y)(t+z)}}.\eqn\labzd
$$
Rotating to the original frame by means of ${\bf r}=Q{\bf r'}$ yields the 
matrix
$$
\Phi_{ell} = 
GM_e\,\left[Q\,{\rm diag}\left(R_D(\Lambda^{22}, 
\Lambda^{33}, \Lambda^{11}), R_D(\Lambda^{11}, \Lambda^{33}, 
\Lambda^{22}), R_D(\Lambda^{11}, \Lambda^{22}, 
\Lambda^{33})\right) Q^T\right].
\eqn\labc
$$

Although the potential inside a homogeneous
ellipsoid is quadratic, the 
potential outside is rather complicated.  This causes the smooth background outside
the ellipsoid to warp, producing a non-quadratic potential inside the ellipsoid
and causing it to become inhomogeneous.  In order to avoid this problem, we assume
that the 
background remains smooth, with a density equal to that of the unperturbed
universe (Icke 1973; White \& Silk 1979). 
The mass of the
ellipsoid, including the contribution from the background density in the volume
covered by the ellipsoid, is taken to be a constant denoted by $M$.  We then compute
the gravitational potential as the sum of two pieces.  The first piece comes 
from the smooth background of density $\rho_b$, yielding
$\Phi_{sph}({\bf r}) = 2\pi G\rho_br^2/3$ and
$$
\Phi_{sph}^{\alpha\beta} = {4\pi\over3}G\rho_b\,I^{\alpha\beta},
\eqn\labs
$$
where $I$ is the identity matrix.
The second piece is associated with the remaining mass of the ellipsoid, 
and is given by
$\Phi_{ell}$ in equation $\labc$ with the mass $M_e\equiv M-\rho_bV$, where
$V = 4\pi\det(\sqrt\Lambda)/3$ is the volume of the ellipsoid.  
This approximation agrees well 
with N-body simulations (White 1993),
because at early times the background is still smooth while at late
times the background has a much lower density than the ellipsoid.

The evolution of the background density $\rho_b\propto a^{-3}$ in 
a matter-dominated universe can be obtained from the FRW equation 
for the scale factor $a=1/(1+z)$,
$$
\left({da\over dt}\right)^2 = \Omega a^{-1} +  \Omega_\Lambda a^2 + \Omega_R.
\eqn\labk
$$
We define
$\Omega = {8\pi G \rho_b / 3 H_0^2}$, $\Omega_\Lambda = {\Lambda_v / 3H_0^2}$,
and $\Omega_R = 1 - \Omega - \Omega_\Lambda$
to quantify the contributions of non-relativistic matter, 
the cosmological constant $\Lambda_v$, and the space curvature to the expansion
of the universe.  These quantities are evaluated at the present time,
and $H_0\equiv d(\ln a)/dt\vert_{a=1}$ is the present Hubble constant. 
 
Next, we note that the $\ell =2$ terms of the external shear [Eq. $\labe$]
also produce a quadratic potential.
Manipulating the spherical harmonics in equation $\labe$
we find
$$
\Phi_{shear} = G\sqrt{\pi\over5}\,\pmatrix{
2\sqrt6{\rm Re\,}a_{22} - 2a_{20}&-2\sqrt6{\rm Im\, }a_{22}&-2\sqrt6{\rm Re\, }a_{21}\cr
-2\sqrt6{\rm Im \,}a_{22}&-2\sqrt6{\rm Re \,}a_{22} - 2a_{20}&2\sqrt6{\rm Im\, }a_{21}\cr
-2\sqrt6{\rm Re\, }a_{21}&2\sqrt6{\rm Im \,}a_{21}&4a_{20}\cr}.
\eqn\labr
$$
Here the relation 
$a_{2(-m)}=(-1)^ma_{2m}^*$ was used to eliminate $m<0$. We choose
the real and imaginary parts of $a_{22}$ and $a_{21}$ as well as $a_{20}$, which
is always real, as the five independent real values in the $\ell=2$ decomposition.
We will discuss the time dependence of the $a_{2m}$ 
coefficients in \S 3 and 
the issue of how to pick them initially in Appendix A.  

With the equation of motion $\labg$ and the potential matrix 
$\Phi^{\alpha\beta}\equiv (\Phi_{sph}^{\alpha\beta} +\Phi_{ell}^{\alpha\beta}
+\Phi_{shear}^{\alpha\beta})$, we may evolve the system of a homogeneous ellipsoid
on a smooth background undergoing a time-dependent quadrupole external shear as
a set of nine second-order differential equations.  
For a non-zero cosmological constant, one should also add:
$\Phi_{vac}^{\alpha\beta}=-{1\over 3}\Lambda_v I^{\alpha\beta}$.
In order to apply this 
approach to a 
collapsing perturbation, we must relate the initial conditions of the ellipsoid
and the time-dependence of the external shear to properties of the initial density
field.  As described above, we center the origin of the coordinate systems
on a high density region and construct a sphere about this point to distinguish
the collapsing object from its environment.
We consider the system at high redshift, so that $\delta\ll1$.
The coefficients $a_{2m}$ that describe the external 
potential as a function of the initial density field are 
given in equation \labf.
We pick the
ellipsoid to match the average density, mass, and quadrupole moments of the inner
spherical region at the initial time. This choice 
is independent of initial time to leading order in the linear regime.
We define the average overdensity of the 
inner region at the initial time as
$$
\overline\delta(R) = 
\left({3\over4\pi R^3}\right) \int_{|{\bf r}|<R}{d^3r\,\delta({\bf r})}.
\eqn\labo
$$
The mass of the region is $M = (4\pi/3)\rho_bR^3[1+\overline\delta(R)]$,
and the quadrupole moments $q_{2m}$ are defined as
$$
q_{2m} = \rho_b\int_{|{\bf r}|<R}{d^3r\,\delta({\bf r})\,r^2Y_{2m}^*}.
\eqn\quadmom
$$
To match these quantities, we consider an ellipsoid with semi-axes $c_1$, $c_2$, 
and $c_3$ oriented at some angle relative to the coordinate axes.
We pick the overdensity of the ellipsoid to be $\overline\delta(R)$.  Then
the mass of the ellipsoid is $(4\pi/3)\rho_b[1+\overline\delta(R)]c_1c_2c_3$,
which when matched to the mass $M$ of the inner region gives 
$$
c_1c_2c_3 = R^3.
\eqn\labh$$
Finally, the quadrupole moments of the ellipsoidal overdensity are chosen to
match those of the actual inner region.
First, we label the points of the ellipsoid by ${\bf r} = QC{\bf x}$, where
${\bf x}$ is a point inside the unit sphere, $C = {\rm diag}(c_1,c_2,c_3)$, and
$Q$ is an orthogonal matrix whose {\it j}th column is the direction of the axis
with length $c_j$.  We then define the matrix $N = QC^2Q^T$ and note the following
integral over the volume of the ellipsoid,
$$\eqalign{\rho_b\overline\delta\int_{ellipsoid}{d^3r\,r_ir_j} 
&= \rho_b\overline\delta c_1c_2c_3\int_{|{\bf x}|\le1}{d^3x(Q_{im}C_{mm}x_m) 
(Q_{jn}C_{nn}x_n)} \cr
&= \rho_b\overline\delta R^3\left({4\pi\delta_{mn}\over15}\right) 
Q_{im}C_{mm} Q_{jn}C_{nn} = {M_e\over5} N_{ij} ,\cr}\eqn\labze
$$
where $M_e = (4\pi/3) \rho_b\overline\delta(R)R^3$ is the mass of the ellipsoid
above the background.
Now we may find the $q_{2m}$ of the ellipsoid in terms of the matrix $N$; for
example, 
$$
{\rm Re}\,q_{22} = \rho_b\overline\delta\int_{ellipsoid}{d^3r \sqrt{15\over32\pi}(r_1^2-r_2^2)} 
= M_e\sqrt{3\over160\pi}(N_{11}-N_{22}).\eqn\labzf
$$
These relations may be inverted, so that given all the $q_{2m}$ we
find $N$ to be
$$
N = \sqrt{40\pi\over3}{1\over M_e}\pmatrix{
{\rm Re\,}q_{22} - {1\over\sqrt6}q_{20}&-{\rm Im\,}q_{22}&-{\rm Re\,}q_{21}\cr
-{\rm Im\,}q_{22}&-{\rm Re\,}q_{22} - {1\over\sqrt6}q_{20}&{\rm Im\,}q_{21}\cr
-{\rm Re\,}q_{21}&{\rm Im\,}q_{21}&{2\over\sqrt6}q_{20}\cr}\;+\;\tau I,
\eqn\labq
$$
where the constant $\tau$ in front of the identity matrix is unknown.
We then diagonalize $N$ to find $Q$ and $C^2$, the latter depending on $\tau$,
and impose the condition in equation $\labh$ to determine $\tau$.
With this condition we find the lengths and directions of the axes of the ellipsoid.
We then set the initial value of $A$ to be $QC$.

Since the equation of motion is second-order, one must specify the 
initial velocities.  We pick the velocities so that the density field is a pure
growing mode, consistent with the fact that we normalize the 
power spectrum today 
when only the growing mode had survived.  
In linear theory the peculiar velocities are given by (Peebles 1980, \S 14)
$$
\delta{\bf v} = {2f{\bf g}\over3H\Omega},\eqn\labzg
$$
where ${\bf g}$ is the peculiar gravitational acceleration,
$H$ is Hubble's constant, and $\Omega=\rho_b/\rho_c$
is the cosmological density parameter.  Here, $f=(a/D)(dD/da)\approx\Omega^{0.6}$,
$a$ is the expansion factor of the universe, and $D$ is the linear
growth factor.  Since ${\bf g} = -\Phi{\bf r} = -\Phi A {\bf x}$, the 
initial velocity field is 
$$
{\bf v} = {dA\over dt}{\bf x} = H{\bf r} - {2f\over 3H\Omega}\Phi{\bf r},\eqn\labzh
$$
and so we find
$$
{dA^{\alpha\beta}\over dt} = HA^{\alpha\beta} - {2f\over 3H\Omega}\Phi^{\alpha\gamma}
A^{\gamma\beta}.
\eqn\labzi
$$
In this way we have related the initial conditions of the equation of motion
to $\overline\delta$ and the five $q_{2m}$.

After evolving the ellipsoid through the matrix $A$, we wish to measure properties
of the ellipsoid, such as its angular momentum and energy.
The axes of the ellipsoid are the eigenvectors of $AA^T$ and the lengths of the 
semi-axes are the square roots of the eigenvalues.  
Labeling these lengths as $c_j$, the potential energy of the
ellipsoid is given by (Binney and Tremaine 1987), 
$$
W = -{3\over5}GM^2 \int_0^\infty{d\tau\over
\sqrt{(\tau+c_1^2)(\tau+c_2^2)(\tau+c_3^2)}}.\eqn\labzj
$$
Here we have neglected the potential energy from the background density and from
the tidal shear, both of which are small at late times
when this quantity is of interest and when $M_e\approx M$.
The kinetic energy equals (Peebles 1980, \S 20),
$$
T = {1\over2}\rho_{e}\int{{\bf v}^2d^3r} = 
{\rho_{e}\det A\over2}\int_{|{\bf x}|<1}{{dA^{ik}\over dt}{dA^{jk}\over dt}x^ix^j}
= {M\over 10}{\,\rm tr}\left({dA\over dt}{dA^T\over dt}\right),\eqn\labzk
$$
where $\rho_{e}$ is the ellipsoid density.
The total energy is $E=T+W$.
Next, the $j$th component of the angular momentum is 
$$
L_j = \rho_{e}\int{({\bf r}\times{\bf v})_jd^3r} =
\rho_{e}\det A\int_{|{\bf x}|<1}{\epsilon^{jkm}A^{kn}{dA^{mp}\over dt}x^nx^p}
= {M\over 5}\epsilon^{jkm}{dA^{mn}\over dt}A^{kn},\eqn\labzl
$$
where $\epsilon^{jkm}$ is the Levi-Civita anti-symmetric tensor.
The mass of the ellipsoid $M$ is constant, and at late times the contribution
from the smooth background is small.
With these definitions we can construct the spin parameter $\lambda$
that measures the amount of rigid rotation acquired by the ellipsoid 
before virialization,
$$
\lambda \equiv {L\sqrt{|E|}\over GM^{5/2}} .\eqn\labzm
$$

\bigskip
\bigskip
\centerline{\bf 3. TIME DEPENDENCE OF THE EXTERNAL SHEAR}

In the equation of motion $\labg$ for the ellipsoid,  
one must specify the time dependence of the external shear.  
The model described in \S 2 considers only quadrupole tidal forces 
and therefore requires a limited amount of information about the  
mass distribution outside the ellipsoid boundary.
In this section, we will try to develop a simple model for the time dependence 
of the $a_{2m}$ coefficients.

The simplest approach is to treat the entire volume outside the spherical
boundary of the inner region by linear perturbation theory.  The motivation
for this approach is that because the inner region is a high density peak, it is likely
to be the first object in its neighborhood to collapse.  Other high peaks
are sufficiently far away that their collapse does not substantially change the
tidal force, since such collapses do not move matter on large angular scales
as seen from the origin.  If the tidal field is determined by scales much
larger than the inner region, the evolution of the quadrupole moments would be
well treated by linear theory.

In linear theory, the fractional overdensity $\delta({\bf x})$ scales by a uniform
growth factor $D(t)$ when considered in comoving coordinates (Peebles 1980).  
>From equation $\labf$, we have 
$$
a_{2m} = \rho_b \int_{|{\bf s}|>R}{d^3s\, Y^*_{2m} \delta({\bf s}) s^{-3}}. 
\eqn\labj
$$
where the ${\bf s}$ coordinates are in physical (non-expanding) space.  Rescaling
to comoving coordinates, however, makes no difference because the magnitude of
$s$ appears equally in the numerator and denominator.
The density $\rho_b$ scales as $(1+z)^3$, where $z$ is redshift.  Thus in linear theory,
$$
a_{2m}(t) = a_{2m}(t_i){D(t)\over D(t_i)}\left({1+z[t]\over 1+z[t_i]}\right)^3. 
\eqn\labi
$$
where $t_i$ is the time when we pick the initial conditions.  In a flat
matter-dominated universe, $D\propto (1+z)^{-1}$ and $a_{2m}\propto (1+z)^2$.

The time dependence given in equation $\labi$ 
can be substituted into the equation of motion for the ellipsoid.
When this is done and an ensemble of random ellipsoid realizations  
are integrated until full collapse, we 
find the average value of the spin parameter $\lambda$ to
be smaller than 0.01.  Since N-body simulations typically find
$\left<\lambda\right> \approx 0.05$
(Barnes \& Efstathiou 1987; Warren \etal 1992), the linear approach apparently
underestimates the external torque.  
It is not surprising that the external shear cannot
be purely described by linear theory, since the
material just outside the boundary of the object has a density similar to that
of the object.  As the ellipsoid collapses, the nearest shell surrounding it follows
its infall.  Since the contribution of a mass element to $a_{2m}$ depends on the inverse
cube of its distance from the origin, the fact that material near the object tends
to be closer than its initial comoving position means that it should produce more torque
than linear theory predicts.  

Based on the fact that the material around the peak tends to be closer than
linear theory predicts, we divide the external 
mass distribution into $N$ spherical shells,
all centered on the origin, and evolve each shell according to the 
mass interior to it (Gunn \& Gott 1972).  
Each shell carries a fraction of the quadrupole shear, initially 
determined from the Gaussian random field, and this shear is scaled according to
the radial collapse of the shell.  The essential approximation is that we neglect
the tangential redistribution of the matter within shells. 
This approximation is justified so long as the motion of the matter 
subtends a small angle, as viewed from the origin, which is always true
for matter at large radius.
Using growing mode velocities, we  
need to specify only the average overdensity interior to the shell
$\overline\delta$ in order to determine its motion.

To calculate the time dependence of the shear in this approximation, we first 
note that $a_{2m}$ depends on $\delta\rho({\bf x})$ but not on rescaling of the radial 
coordinate.  We find $\delta\rho$ by solving the spherical evolution
from some initial overdensity $\overline\delta(t_i)$ using the FRW equation $\labk$, 
$$
{d^2r\over dt^2} = -{4\pi G \rho_i\over 3r^2} + {\Lambda_v r\over 3},
\eqn\labm
$$
with initial density $\rho_i=(1+\overline\delta)\rho_b(t_i)$ and $r(t_i)=1$.  
The initial velocity is to leading order 
$H(t_i)(1-\overline\delta/3),$ with
$$
H(t_i) = H_0 \sqrt{\Omega(1+z_i)^3 + \Omega_{R}(1+z_i)^2 + \Omega_\Lambda},\eqn\labzn
$$
where $z_i$ is the redshift at the initial time $t_i$.
Thus for a given $\overline\delta$, we may integrate equation $\labm$
to find $r(t)$.  We then
find $a_{2m}(t)\propto \delta\rho$ by considering the difference between the top-hat
density, $\rho_{sph}\propto r^{-3}$, and the background density, $\rho_b\propto (1+z)^3$.  
Thus,
$$
{a_{2m}(t)\over a_{2m}(t_i)} = 
{(1+\overline\delta_i)r(t)^{-3} - 
\left({[1+z(t)]/[1+z_i]}\right)^3\over \overline\delta_i}.
\eqn\labl
$$
For a sufficiently early initial time in the matter-dominated
epoch, $\overline\delta_i\ll 1$ and $\Omega\approx1$, 
so that equation $\labl$ matches the linear theory result of $a_{2m}\propto (1+z)^2$.

To implement this approach, we wish to consider $N$ shells, with radial 
boundaries $R_\alpha$, for
$\alpha= 0, 1, \ldots, N$.  We take $R_0$ to be the radius of the spherical 
boundary that separates the ellipsoid region from the external region and 
$R_N = \infty$ to indicate that the outer shell includes everything outside the
outermost spherical shell.  Initially each shell carries a fraction of the 
quadrupole shear
$$a^{(n)}_{2m} = \rho_b \int_{R_{n-1}<|{\bf s}|<R_n}{d^3s\, Y^*_{2m} \delta({\bf s}) s^{-3}},\eqn\labn$$
where $n=1,2,\ldots,N$ labels the shell.
We pick the characteristic overdensity of each shell as
$$
\overline\delta^{(n)} \equiv {1\over2}(\overline\delta[R_{n-1}] 
+ \overline\delta[R_n]) .
\eqn\labp
$$
For thin shells the differences between the definitions $\overline\delta^{(n)}$ 
and $\overline\delta(R_n)$ defined in equation $\labo$ is small.
The overdensity surrounding the outermost shell $R_{N-1}$ 
is negligible, 
so we set $\overline\delta^{(N)}= 0$ and treat the
tide of the outermost shell using linear theory.

The above model for the exterior region completes the necessary 
set of equations and allows us to 
evolve the ellipsoid beyond the linear regime according to equation $\labg$.  
This approach requires a large set of 
initial data: the overdensities $\overline\delta$ for all the shells 
and for the inner region, the quadrupole shears
$a^\alpha_{2m}$ for all outer shells, and the quadrupole moments $q_{2m}$ for the inner region.  
This set comprises $6N+5$ real numbers, randomly drawn in a highly correlated
way from a Gaussian random field of initial density perturbations.
By treating the random density field in spherical coordinates, as suggested by Binney and 
Quinn (1991), we can separate this set of random numbers into six independent
sets, each of which is distributed as a multi-dimensional Gaussian distribution.
We describe how to find these distributions in Appendix A.
We may then draw the initial conditions from the proper
probability distribution by taking linear combinations of $6N+5$ random 
Gaussian deviates.  

\bigskip
\bigskip
\centerline{\bf 4. COLLAPSE AND CONSTRAINTS}

If the ellipsoid is simply evolved forward in time, one soon reaches the obvious
problem that the axes collapse at different times.  Typically, one
axis turns shorter and collapses first, forming a pancake 
(Lin, Mestel, Shu 1965; Zel'dovich 1970; Peebles 1980)
in which the baryons shock and the dark matter goes through  
violent relaxation. 
The computation
of angular momentum should not end at the time of the short axis collapse,
since the other two axes
are still extended and give the object a significant quadrupole moment with which
it couples to the tidal shear.  Previous work on spherical objects (Ryden
1988) has shown that most of the angular momentum is acquired near turnaround,
since most of the time is spent at this phase.  In the
ellipsoid model, turnaround is not a single epoch and it appears necessary to
follow the system as long as the quadrupole moments are large, i.e. until
the long axis turns around.  After the short axis
collapses, it makes a small contribution to the quadrupole moment
of the ellipsoid.
The quadrupoles are proportional to the difference between the squares of
the axis lengths and so the exact dynamics
of the virialization process along the short axis 
has little effect on the acquisition of angular momentum.

We therefore follow the approximation of Bond and Myers (1994) and impose a 
strict cutoff on the collapse of an axis, namely that no axis may
collapse below 40\% of its maximum length.  This keeps the dynamics from approaching
the singularity at zero length and simulates the virialization of the corresponding axis.
The 40\% cutoff was picked to allow the ellipsoid to be slightly flatter than
the spherical virial theorem would predict, but because the collapsed axis is much
shorter than the non-collapsed ones, the exact cutoff value does not affect the angular
momentum acquisition significantly.  
The implementation of this constraint is non-trivial
because the ellipsoid may be rotating
and the columns of the matrix $A$ are in general not orthogonal.
The details of this implementation are presented 
in Appendix B.  Note that the vorticity of the ellipsoid
is conserved under the equation of motion $\labg$, but is not conserved
through this treatment of the collapse of an axis.  Also, since
the kinetic energy of the ellipsoid is being altered, energy is not conserved.
Because the change in the total energy during the 
dynamics is usually small, we save
the last value of the energy before the first axis collapses
and use it in the calculation of the
spin parameter $\lambda$.

In principle, we would like to evolve the ellipsoid until all three axes collapse, 
by which time the angular momentum gained by tidal torques (i.e. not by
accretion) will be complete.  The first difficulty is that  
in some cases one or two axes
do not collapse at all.  The existence a high-density region does not
guarantee that it will not be sheared by its environment
to form a wide pancake.
This would be true even if we were to use linear theory
for the shear, as can be seen from the Zel'dovich approximation (Zel'dovich 1970).   
In this approximation, if a volume element begins
with an outward peculiar velocity along one of its axes
then it will continue to move outwards for all
later times.  Because the region is overdense, all the growing mode velocities resulting 
from the ellipsoidal perturbation will be inward.  But since the trace of
$\Phi_{shear}$ is zero at least one eigenvalue of $\Phi_{shear}$ is negative,
and therefore in at least one direction, the peculiar velocity will be less
inward than for the isolated ellipsoid.  We wish to study regions that end as
spatially bounded objects.  In $\Omega=1$ cosmologies, following the Zel'dovich
approximation, we require that the radial component of the initial 
peculiar velocities inward in all directions.  This corresponds to requiring 
that the potential matrix be positive definite.  In $\Omega<1$ cosmologies,
however, this condition is not sufficient because mildly overdense regions
may still expand forever.  Only regions which exceed the critical
density collapse, so we instead require that the inward radial component
of the peculiar velocities exceed that of the growing mode of 
a spherical top-hat perturbation at the critical
density, which corresponds to an initial overdensity 
$$\overline\delta_{crit} = {3\over5}\left({\Omega_R\over\Omega}(1+z_i)^{-1} +
{\Omega_\Lambda\over\Omega}(1+z_i)^{-3}\right).\eqn\labw$$
Due to nonlinear effects of the 
external shear, our requirement is not a perfect discriminator
as to whether an axis will expand forever,
but it at least matches the weakly non-linear theory.  The
consequences of this constraint are examined analytically in Appendix C.

Since we base the time dependence of the shear on
the density of the spherical shell collapse, the shear force diverges
as the shell collapses to a zero radius.  Because the average overdensity of the 
innermost shell is not so different from that of the ellipsoid, the two objects
collapse at similar times.  But in most cases, the action of the shear
causes the long axis to collapse significantly later than the inner external shell.  
We must end the
integration at some time before the collapse of the inner tidal shell.
We therefore choose to end at the time at which a sphere whose overdensity was
that of the initial ellipsoid would collapse to a zero radius.  We then constrain
the initial conditions so that none of the exterior 
shells has an overdensity $\overline\delta^{(n)}$ greater than
95\% of the initial density of the ellipsoid.  This ensures that the external tidal
shells do not collapse before the integration ends.
This condition has the important property of tending to
make the inner region a density peak; we do not explicitly require that it is
truly a peak, but if the overdensity as a function of radius is forced to drop
as one moves away from the central region, only extreme variations as a function of
angle would allow the density to rise radially in a particular direction.

The last and most significant
constraint we impose is that the inner region must have a high overdensity. This condition 
underlines the separation 
between the collapsing region and the rest of the universe.
We implement
this constraint 
by requiring $\overline\delta(R_0)>\nu_{min}\sigma$,
where $\nu_{min}=const$ and $\sigma$ is the {\it r.m.s.} amplitude
of mass fluctuations $\delta M/M$ for such regions.
As usual, the latter is calculated to be (Bardeen \etal 1986),
$$
\sigma^2 = {1\over 2\pi^2} \int^\infty_0{dk\,P(k) \left({3j_1(kR_0)\over R_0}\right)^2},\eqn\labv
$$
where $P(k)$ is the power spectrum and $j_1$ is a spherical Bessel function.
In \S 5 we will typically pick $\nu_{min}\gsim2$, 
thus forcing the central region to be rare.

The above three constraints---(i) $\overline\delta(R_0)>\nu_{min}\sigma$;
(ii) the overdensity of the surrounding shells being less than 95\% of
$\overline\delta(R_0)$;
and (iii) all peculiar velocities being inwards---are not implemented directly
into the probability 
distributions for the initial conditions as derived in Appendix A.  Rather,
we generate sets of the initial conditions and then reject cases that 
do not satisfy these constraints.  Since
the strictest condition is $\overline\delta(R_0)>\nu_{min}\sigma$, we set up
the probability distribution so that $\overline\delta(R_0)$ depends on only one
Gaussian deviate, which allows us to reject a set of initial conditions on the
basis of one random number.  Then we can choose the overdensity of the innermost 
shell to depend on only one more deviate, allowing another straightforward test.

We call the fraction of sets of initial data that meet the above conditions the
acceptance rate $A$.
Prior to imposing these constraints, the Gaussian random field was unconstrained.
Because the central region is at a mass scale $M = (4\pi/3)R_0^3$, the number
density of such regions is $n_0 = \rho_0/M$, where $\rho_0$ is the mass density
today of non-relativistic matter (i.e. the matter making up the ellipsoid).
This means that after applying the constraints, the number density of accepted regions is 
$An_0$.
If we now construct an ensemble of such regions and, after evolving each of
them with our model, find that a given property occurs in some fraction $f$ of them,
then the actual number density of such objects is predicted to be $fAn_0$.
We apply this approach elsewhere to predict the number density of black holes form in the initial
collapse of objects with very low values of angular momentum (Eisenstein and Loeb 1994, hereafter 
Paper II).

This completes the description of our model.  We next study the statistical properties 
of collapsing regions in specific
cosmological models, defined by the values of the 
cosmological parameters: $\Omega$, $\Lambda_v$,
and $H_0$, as well as by the power spectrum of primordial density perturbations.  
For a given mass scale, we 
consider many realizations of the random initial conditions, evolve each realization separately, 
and then combine the results to 
analyse the statistical distribution of shapes and angular momenta
of collapsing regions.

\bigskip
\bigskip
\centerline{\bf 5. RESULTS}

To illustrate the behaviour of the model, we adopt a cold dark matter (CDM) power
spectrum (Bardeen \etal 1986) of initial density perturbations 
for two different universes: 
a flat universe with $\Omega =1$,
and an open universe with $\Omega=0.2$;
both cases assuming
$\Lambda_v = 0$ and $H_0 = 50{\,\rm km/s/Mpc}$.
We normalize the power spectrum by choosing the present {\it r.m.s.} 
amplitude of mass fluctuations within an $8h^{-1}{\rm \,Mpc}$ radius sphere to be $1/b$, 
where $b$ is the bias parameter.  We pick the initial time to correspond to redshift 
$z_i = 1000$, and then scale the power 
spectrum using the growth factor $D(t_0)/D(t_i)$.

The parameters of the model are the mass of the 
central region $M$, the minimum peak height $\nu_{min}$, 
and the radii of the external shells.  
We consider
a variety of mass scales between $10^8$ and $10^{15}{\,\rm M_\odot}$.
As peak thresholds we use both $\nu_{min}=2.5$ and 2.0.  
The radius of the innermost spherical
boundary $R_0$ is fixed to match the mass scale being studied, and we pick all
other radii to be multiples of this inner radius.  We choose twenty 
shells, with boundaries at 1.5, 1.75, 2, 2.5, 3, 3.5, 4, 4.5, 5, 6, 7, 8, 9, 10, 12, 
15, 17, 20, and 30 times the inner radius.  The only important number in this
list is the first one, which implied that
the innermost shell is thick.  This is done in order to keep the
acceptance rate reasonable, as the overdensity of the closest outer shell 
must be less than 
95\% of that of the inner region.  Were this shell to be thin, the overdensity constraint
would only rarely be satisfied.  
The finite thickness of this shell ultimately
limits the radial resolution of the external shear profile.

We consider the flat universe with $b=1$ for a variety of mass scales
and for $\nu_{min} = 2.5$.
The average value of $\lambda$ increases slightly with mass, from 0.034
at $10^8{\,\rm M_\odot}$ to 0.038 
at $10^{12}{\,\rm M_\odot}$ and to 0.049
at $10^{15}{\,\rm M_\odot}$.
Figure 1 shows the probability distribution of the spin parameter 
$P(\lambda)$
for the mass scales of $10^{12}{\,\rm M_\odot}$ and 
$10^{15}{\,\rm M_\odot}$ in a flat universe.
The histograms are based on samples of random realizations
that result in $2\times10^5$ accepted systems for each mass scale.  
The weak dependence of the mean values of $\langle \lambda\rangle$
on the mass scale is in good agreement with with N-body simulations
(Barnes \& Efstathiou 1987; Warren et al. 1992). 

The primary reason for the actual dependence of $\langle \lambda\rangle$ on $M$
is the shape of the power spectrum.
With the cold dark matter power spectrum, higher mass scales have larger 
average quadrupole moments
(relative to $\sigma MR^2$).
This results in more elongated initial ellipsoids and 
therefore in stronger couplings to the external torque.
In addition, the fixed choice of shell boundary radii introduces a systematic mass
dependence because the higher mass systems have a steeper density profile 
and a larger fraction of the 
shear carried by the inner shells (these 
two profiles are related, as shown in Appendix C). 
Consequently, the effective shell resolution appears
coarser 
for high mass objects.
The limited resolution results in
a slight
underestimate of the shear at late times, when
the shear amplitude is large.  

For $\nu_{min}=2.5$ only a fraction 0.00621 of all regions
satisfy the constraint $\overline\delta(R_0)>\nu_{min}\sigma$.  The 
remaining two constraints (shell densities and inward velocities) are satisfied 
64.2\% of the time for $10^8{\,\rm M_\odot}$ and 89.7\% 
for $10^{12}{\,\rm M_\odot}$.  Most of these additional rejections are due to the
inner shell having too high an overdensity; rejections due to the peculiar
velocities not being inward occur in less than 3\% of the cases (cf. Appendix C).
With peak threshold values of $\nu_{min}=2.5$ and 2.0, 
the average values of $\nu$ are 2.84 and 2.40 respectively,
independent of mass scale.

For a flat universe, the model is strictly independent of the bias parameter or
the Hubble constant.  Both parameters rescale space and time without
altering the collapse dynamics.
This does not hold in an open universe, where an intrinsic time
scale appears at the transition point 
to an open curvature-dominated expansion of the background.

The value of $\lambda$ is found to be
anti-correlated with $\nu = \overline\delta/\sigma$, so that
higher peaks tend to have lower angular momentum.  The linear correlation
between the two is about 0.21.  Therefore, the choice of $\nu_{min}$ affects
the distribution of $\lambda$. For a flat universe and 
$M=10^{12}{\,\rm M_\odot}$ we find $\left<\lambda\right> = 0.038$ for $\nu_{min} = 2.5$
and $\left<\lambda\right> = 0.051$ for $\nu_{min} = 2.0$.
For $10^{15}{\,\rm M_\odot}$ we find $\left<\lambda\right> = 0.049$ for
$\nu_{min} = 2.5$ and $\left<\lambda\right> = 0.069$ for $\nu_{min} = 2.0$.
If we use a particular $\nu$ for all iterations, 
as opposed to the usual requirement of $\nu>\nu_{min}$, then 
for $10^{15}{\,\rm M_\odot}$ we find $\left<\lambda\right> = 0.089$ for
$\nu=2$, $\left<\lambda\right> = 0.060$ for $\nu=2.5$, and 
$\left<\lambda\right> = 0.042$ for $\nu=3$. 

In all of the accepted objects, the shortest axis reaches collapse by the 
end of the integration
(i.e. by the time a spherical perturbation with 
the same initial density as the ellipsoid would have reached zero radius).  For
$\nu>\nu_{min}=2.5$, the middle axis reaches collapse in 78\% of all iterations, 
while the long axis reaches
turn-around in 84\% of the iterations and reaches 
full collapse in 1.5\% of all cases.  
If we force $\nu=2$, the corresponding numbers are 65\%, 58\%, and 0.6\% 
respectively for 
$M=10^{12}{\,\rm M_\odot}$, with a weak dependence on mass.
The failure of the long axis to collapse is expected; since it is
being pulled out by the shear, it collapses later than the unperturbed spherical
case.  The few times that the long axis happens to collapse result 
from the small window of
opportunity between the sphere reaching 40\% of its turn-around radius (defined
as axis collapse) and reaching a zero radius.  The failure to turn-around usually
is the result of the shear leaving one axis with only a tiny inward velocity so that
it undergoes a very long excursion, but it can also result from extreme cases
where the rapidly increasing shear toward the end of the integration causes
one axis to expand again after it had already been contracting.  
The dependence of the amount of collapse on $\nu$ reflects 
the tendency of the shear to pull lower peaks apart. Although we require 
that the velocities are initially inwards, this may not be sufficient
to guarantee that the object remains bounded under the influence of the nonlinear
shear.  As suggested in Appendix C, the constraint to have inward velocities 
is almost negligible for $\nu>2.5$, with acceptances over 97\%, but is important
for $\nu=2$, where the acceptance drops to 82\%.  The drop means that more systems
have one axis with a nearly zero peculiar velocity; these axes are very
slow to turn-around. Low-$\nu$ systems therefore 
have larger axis ratios resulting in larger couplings to 
the quadrupole torques. The correlation between the peak height $\nu$ and the
axis ratios leads to the above dependence of $\lambda$ on $\nu$.
 
Let us now turn to discuss the axis ratios of the evolving ellipsoids.  Figures 2-8 
present the axis ratios $b/a$ and $c/a$ for various ensembles of ellipsoids.
The semi-axis lengths $a$, $b$, and $c$ are ordered
so that $a\geq b\geq c$.  Panels (a)-(e) show the ensemble at different 
times, while panel (f) shows the mean of the points in the previous panels.

Figure 2 presents the evolution of the axis ratios for a random ensemble of 
$10^{15}{\,\rm M_\odot}$ ellipsoids with $\nu=2$ in the $\Omega=1$ cosmology.
We chose to fix $\nu$ rather than use a threshold value $\nu_{min}$ so
that the integration of each ellipsoid would end at the same time.
The time slices in the plots show each ellipsoid at a constant fraction of the
collapse time  $t_{\rm max}$ for a spherical top-hat perturbation 
of the same initial density.  We track
5000 ellipsoids and show the axis ratios at the initial time and 25\%,
50\%, 75\%, and 100\% of $t_{\rm max}$.  We
set $b=1.3$ so that the end time is close to $z=0$.

The most significant aspect of this plot is the progression of the points from a 
quasi-spherical initial state to a prolate end state. At early times [panels (a)-(c)]
the short axis falls behind in the expansion 
so that the points move down the plot.  At the time of panel (d) the short
axis has generally collapsed; however
the typical resulting shape is not an oblate
pancake or a prolate filament but rather triaxial.  
As the evolution continues the middle
axis turns around and collapses, so that most objects become prolate
by the end time in panel (e).  Note the slight regeneration of high $c/a$ points
between panels (d) and (e); this corresponds to the beginning of a collapse along
the long axis.  Indeed, 37 of these 5000 ellipsoids reached long-axis collapse 
(i.e. the long axis reached 40\% of its maximum length).

In contrast to previous discussions on collapsing ellipsoids 
(Lin, Mestel \& Shu 1965; Zel'dovich 1965; Icke 1973; White \& Silk 1979),
we find that the collapsing region geometry is primarily determined by the 
external shear and not by the initial anisotropy of the ellipsoids. 
To demonstrate this result, we alter the model
and replace the initial ellipsoid by a sphere. The results
for a mass of $10^{15}{\,\rm M_\odot}$, $\Omega=1$, and $\nu=2$, are
shown in Figure 3.  
The evolution is very similar 
to Figure 2 despite the qualitative change in the initial conditions,
although the ratio of the long axis to the short axis is not quite 
as large at late times. 
Furthermore, even in the full ellipsoid model,
the direction of the long axis
is dominated by the direction of the shear rather than the direction of
the long axis of the initial ellipsoid.  This is particularly the case
for low mass scales and for higher $\nu$, where the initial quadrupoles are
weak and the initial conditions are closer to the spherical example.  
Even an initially spherical object can
gain angular momentum because the shear direction changes as the relative
weighting of the shells evolves with time.
The resulting value 
of $\lambda$ is  lower for the spherical case, however, 
indicating that the anisotropy of the 
initial ellipsoid is important for the acquisition of angular momentum.
For the parameters in this paragraph, 
$\left<\lambda\right> = 0.089$ when the initial object is the usual ellipsoid,
but $\left<\lambda\right> = 0.028$ when the initial object is changed to a
sphere.

In Figure 4, we consider the mass scale of $10^{12}{\,\rm M_\odot}$
and $\nu=2$ in 
the $\Omega=1$ cosmology (the redshifts assume $b=1.3$).  
Compared to Figure 2, the 
initial ellipsoids are more spherical 
and have weaker quadrupole moments.
However, the axis ratios at
late times are similar to those at 
the $10^{15}{\,\rm M_\odot}$ mass scale. 

We next consider the open cosmology with $\Omega = 0.2$ and $b=1$.
For the $10^{15}{\,\rm M_\odot}$ mass scale we pick $\nu = 2.2$ so that the
final redshift of the calculation matches that of the flat universe case.  The results 
are presented in Figure 5. Here we have chosen the time slices so that
the redshifts are the same as in Figure 2.
The slices are no longer evenly spaced in time
because of the different cosmology.
Due to changes in the power spectrum,
the objects are initially more spherical in an open universe,
but they end with very similar axis ratios to those 
obtained in a flat cosmology.

A major difference between the open and the flat cases is that the 
constraint on the initial velocities is far more severe in the
open cosmology. In the open case we require that the peculiar velocities 
not only be inward but that they be larger in magnitude than those induced
by a critically bound sphere.
For the above open cosmology and the above mass scale 
this condition [cf. Eq. $\labw$] requires that a 
spherical top-hat perturbation have $\nu>1.1$.
Including the disruptive effects of shear
(cf. Appendix C), we find that the acceptance rate for this
constraint drops down to 23\% from its value of 83\% for $\Omega=1$.  
Thus, most $\nu=2$ peaks have one axis that is not critically bound.
In Figure 6, we relax the constraint on the velocities entirely so as to investigate
how the general peak behaves in this open universe.  There are two notable
differences between this plot and either Figure 5 
or Figure 2.  First, the collapse of the shortest axis,
as shown by the accumulation of the points near the horizontal axis,
occurs in panel (c) rather than in panel (d).  The relaxation of the velocity 
constraint allows larger shears to enter. This not only tends to
pull out the long axis but also tends to push in the short axis, since
the shear is a traceless matrix. The short axis in this sample therefore collapses
faster.  Second, at the final time in 
panel (e), the objects are significantly more filamentary (note
that small deviations in the plotted ratio $c/a$ near zero 
correspond to large changes in the filament shape $a/c$).  
This effect occurs because the long axis expands without bound
and the short axis collapses earlier than before.
Comparing Figures 2 and 6, we see that when viewed at equal redshifts, open 
universes have significantly more filamentary structure than 
flat universes. This occurs primarily because
of the stronger influence of shear on the collapsing regions.

Next, we consider the low-spin tail of the population of collapsing regions
(cf. Fig. 1). Because of its low
angular momentum, the gaseous component of 
the systems in this tail can collapse to small radii 
and form compact massive objects. The low-spin systems therefore provide
environments that may favor the formation of 
seeds for quasar black holes; we discuss this possibility
in detail in Paper II.
For the present discussion, let us examine the qualitative
properties of low-spin systems selected out of a population of
collapsing regions with $\nu_{min}=2.5$ on the mass scale of
$10^{8}{\,\rm M_\odot}$ in a flat universe (including the velocity constraint as usual).  
Figure 7 shows the full distribution 
of the ensemble of runs for this mass scale.
The different panels show the ellipsoid axis ratios
at fixed fractions of the collapse time $t_{\rm max}$
of a spherical perturbation with the same initial density
as the ellipsoid.  
Since $\nu$ is not constant,
the collapse time is different for different ellipsoids. Thus, the 
panels in the figure are not equal time slices, but instead show each ellipsoid relative 
to a measure of time scale for its development.

We now compare this full distribution to 
the distribution of the subset of low-spin systems. 
For the definition of a low-spin system, we use the criterion 
developed in Paper II to determine whether the gas in a
system can settle into a sufficiently compact disk with a viscous
timescale shorter than the star formation time $\sim 10^7\,\rm yrs$ (i.e. 
$y<1.1\times10^{-3}$ in the notation of Paper II).
This condition translates to 
a maximum value for the angular momentum per unit mass of the system,   
$J/M = 8.6\times 10^{24}\,{\rm \,cm^2\,s^{-1}}$. Since this threshold involves 
only the total angular momentum, it does not precisely correspond 
to a bound on the spin-parameter $\lambda$ 
(which depends on the energy as well); nevertheless, it is approximately the 
condition $\lambda\lsim 10^{-3}$.  
In a set of $2\times10^5$ accepted systems,
232 satisfy this limit.  In Figure 8, we plot the axis ratio histories of these objects.
Evidently, these objects evolve in a much more spherical way than the typical cases
shown in Figure 7; over half of the low-spin systems have axis ratios 
between 2:1 and 1:1 at the end time. This striking result is a consequence
of two correlated effects. First, the low-spin systems have closer to spherical
initial conditions, but second and more important, the quadrupole 
shear in their neighborhood is weak. 
One should also consider the existence of
a background population of accidental low-spin
objects.  Because of the time dependence of the shear, it is possible for the angular momentum
of a filamentary object to be rapidly changing at late times and 
by chance be small at the end of the integration.  Such elongated objects are obviously 
not environments that would favor black hole formation.
While some of these background objects do occur, 
Figure 8 demonstrates that most
of the low-spin events are indeed close to spherical 
during their evolution.

We close this section by showing four individual cases for
the collapse of a $10^{15}{\,\rm M_\odot}$ object 
in an $\Omega=1$ universe with $b=1.3$.  
Figures 9a and 9b show two objects with $\nu=2$, while Figures 9c and 9d
show two objects with $\nu=3$.  In all cases, the first panel shows the
semi-axis lengths as a function of time (note the different time scales between
the two values of $\nu$).  The freezing of the axis length after its 
collapse (cf. Appendix B) is evident.  The second panel shows the evolution of the 
overdensity $\delta\rho/\rho = \overline\delta$ for the ellipsoid (solid line), as well
as for the spherical perturbation with the same initial density (dotted)
and for the Zel'dovich approximation when applied to the initial quadratic potential
(dashed).  The slope discontinuities in the ellipsoid density are a result
of halting the collapse of the axes; if this treatment was avoided 
the ellipsoid would have reached an infinite density slightly before its
spherical counterpart.
The Zel'dovich approximation does well at early times but underestimates the
density after turn-around.  The third panel shows
the spin parameter as a function of time.  While $\lambda$ grows roughly linearly
with time, there are significant secondary variations which are more pronounced at lower
mass scales.
The four different figures are shown to illustrate a variety of collapse conditions. 
Figure 9a ends with relatively large values of $\lambda$ and the axis ratios.
Figure 9b was picked because of the unusual wiggle in its evolution; such 
wiggles result from a time-varying shear and are not uncommon in the 
$\nu=2$ systems when the overdensity is not high enough to dominate the 
shear.
Figure 9c is a typical high-$\nu$ object; its $\lambda$ and axis ratios
are close to the mean. Finally, Figure 9d has a relatively 
low $\lambda$ but a large ratio of its long to short axes.

\bigskip
\bigskip
\centerline{\bf 6. DISCUSSION AND CONCLUSIONS}

In this work we have developed a model for the non-linear collapse of 
a triaxial overdense region out of a Gaussian random field of primordial
density perturbations. The model approximates the collapsing region as 
a homogeneous ellipsoid. We assume that the collapsing mass 
originates in a spherical volume
around a high density peak\footnote{2}{As 
shown in Appendix C, only high density peaks
are likely to survive as bound systems under the influence
of the external tidal shear from their environment. 
This effect provides another reason for associating virialized
objects, like galaxies or clusters, with high density peaks.}
and select an ellipsoid that matches
the mass, mean overdensity, and quadrupole moment of the
initial overdensity field
in this volume. This choice 
is independent of redshift to leading order 
in linear perturbation theory. The mass distribution 
outside the sphere exerts a tidal torque on the 
ellipsoid and spins it up. 
We calculate
the quadrupole moment of the external shear 
as a function of time by dividing the 
background density field into thin spherical shells
that move only radially
according to the mass interior to them.
The dynamics of the ellipsoid is determined by its self-gravity 
and the external shear through a set of nine
ordinary differential equations. Both forces are linear in the 
coordinates and therefore maintain
homogeneity of the ellipsoid at all times. 
In Appendix A we have developed the formalism necessary to 
randomly determine the initial conditions for this model in the appropriate correlated
way from a Gaussian random field of initial density perturbations.

The above model was applied to a restricted set of initial conditions
that are more suitable to its assumptions. In particular, we studied 
the statistical properties of rare
high density peaks with a mean overdensity $\overline{\delta}\gsim2\sigma$.
In a bottom-up hierarchy of structure 
formation, most objects evolve 
from a quasi-spherical initial state
to a pancake or a filament and then to complete virialization.
As demonstrated by Figure 3 (where the initial conditions are spherical),
this evolution history of shapes is primarily induced by the 
tidal shear and not by the initial triaxiality of the ellipsoids.
Thus, {\it the existence of sheets and filaments in the universe is not a result
of the Lin-Mestel-Shu (1965) instability}, but rather an
environmental effect; namely the ellipsoids are being sheared by nearby 
mass concentrations. As shown in Figures 2 and 6, the redshift evolution of the
triaxiality of systems on a given mass scale
can be used to discriminate between
an open and a flat universe.
However, the average value of the spin parameter
(cf. Fig. 1) $\langle\lambda\rangle\approx 0.04$,
is found to be only
weakly dependent on the object mass or the cosmological parameters,
in agreement with N-body simulations.  There is a modest dependence of
$\lambda$ on the peak height $\nu$.

The ellipsoid model incorporates significant qualitative improvements over previous analytical
investigations of $\lambda$.  Other models (Ryden 1988; Quinn \& Binney 1992)
considered spherical dynamics for the collapsing region
and assumed zero initial peculiar velocities rather than 
growing mode velocities. The latter assumption causes a 
significant overestimate of $\lambda$ 
since it makes the object expand to a larger radius.  
Quinn \& Binney (1992) considered the dipole term of the
external potential to be most important, but they
worked in the instantaneous rest frame of the center of mass rather than in
the actual accelerating rest frame of the center of mass.
This led to a 
nonzero torque
even in a uniform gravitational field.

In assessing the applicability of our model, we return to 
the initial ellipsoid. The matter inside this ellipsoid makes up the final object.
The ellipsoid is picked to match the mean overdensity
and quadrupole moment of the inner spherical region.
As it traces the density, one may consider the boundary of the ellipsoid
as an approximation to the smoothed isodensity contour of the inner region.  The 
model then follows the evolution of
this contour and would therefore be useful for applications where the final object
is indeed associated with the initial high-density contours.

A potential example for such an application is the collapse
of clusters of galaxies.  
Clusters correspond to high-density peaks on the
$10^{14-15}\,M_\odot$ mass scale.  
In the bottom-up hierarchy of structure
formation the galaxies that inhabit the cluster form before
the cluster collapses.  Since galaxies tend to form in high-density regions, the 
high-density environment of the cluster peak favors galaxy formation. 
Galaxies therefore tend to trace the smoothed density field near the top of the peak,
and the ellipsoid model could describe their motion
during the formation of the cluster. Obviously the model cannot describe
the final virialization process
of clusters, where violent relaxation erases in part the signature of
the initial conditions. The ellipsoid model 
predicts however that the outer unvirialized parts of clusters
which are still infalling today would be highly triaxial 
with the axes ratio distribution as shown in Figures 2 and 5.
This prediction can be tested by observations of the 
galaxy distribution around clusters 
or by deep x-ray imaging of their surrounding gas distribution. 
In fact, the Virgo cluster is observed to be elongated considerably 
along the line of sight based on Tully-Fisher distances (Fukugita, Okamura, \& Yasuda 1993). 
The existence of
prolate systems allows for a systematic contamination of optical samples of rich 
clusters by cases of a chance alignment between a prolate filament
and the line of sight.
Similarly, clusters that are elongated perpendicular to the line of sight
may not have been identified as clusters when examined by spherical filters 
(e.g. Abell 1958).
In addition, prolate clusters introduce 
a systematic bias into the Sunyaev-Zel'dovich (SZ) effect
by favoring low values of the Hubble constant when the data analysis is done  
using a spherical model for the cluster. Indeed, attempts 
to determine the Hubble constant from SZ measurements tend to get relatively
low values of $H_0$, 
occasionally below the permitted lower 
bound of $50~ {\rm km~s^{-1} Mpc^{-1}}$ (McHardy et al. 
1990; Birkinshaw \& Hughes 1994, and references
therein). Conversely, when the value of the Hubble constant is
eventually determined by other techniques,
it would be possible to get constraints on the triaxiality 
of clusters from their SZ effect.

In some other applications, the final object 
evolves from a region that has
little to do with the density contours near the top of a high density peak.  
For example, the matter that forms a galactic halo 
may come from regions of low initial overdensity near the peak.  
These low density regions can be assembled into the halo not only by the
gravitational attraction of the peak 
itself but also by the external shear.  Initially, 
the total densities of the ``low'' and ``high'' density regions are similar
because the background density dominates in both cases. The actual boundary
of a virialized object today is an equal collapse-time surface that maps into 
a complicated surface in the initial density field according to
the intervening action of the external shear.
To illustrate this complicated situation, consider a high overdensity region
acted upon by a quadrupole tidal shear.  The initial region has a slightly anisotropic
shape, but this is quickly counteracted by the shear (cf. Figs. 2 and 3).  
The density contours
are pulled into a strongly triaxial shape, as one axis is pushed in by the shear
and a second axis is pulled out; the third axis is somewhere in between.
As the central peak collapses along its short axis,
which is more likely to be considered as part of the object: the high density regions
located far out on the long axis or the lower density regions located nearby
on the short axis? For the formation of a galactic halo, the nearby regions are more relevant, 
since they are positioned closer to the center of mass 
and therefore undergo shell crossing earlier.
Moreover, the sections of the high initial density contours out on the long
axis are susceptible to being separated from the halo by fragmentation
or other instabilities (Merritt \& Hernquist 1991).
Thus, the initial isodensity contours  
around the peak may not be appropriate tracers
of the volume that eventually makes up the 
halo.

An additional problem is that while a galaxy corresponds to the region surrounding
a high-density peak, the peak itself collapses at high redshifts and the
surrounding matter accretes onto this seed.  Hence, the properties of the
collapsed peak do not reflect the properties of the final halo.  In a spherical
collapse model, one hopes to remedy this problem by picking the radius of the object so
that the overdensity inside that radius corresponds
to a collapse at present.  However, when the effects of shear and 
non-sphericity are included, this procedure is no longer valid.  

For the above reasons, we feel that the ellipsoid model does not provide a
fully satisfactory description of the dynamics of typical galactic halos.
However, the model does substantially better in describing low-spin objects
at high redshifts (cf. Paper II). Such objects are generally 
located in regions with low shear, so that their collapse is more spherical,
as demonstrated 
in Figure 8.  This fact tends to reduce the concerns about
the proper initial volume for the system.  The population of low-spin 
objects is of particular interest as potential environments
that favor the formation 
of massive black holes, since in these systems
the hydrodynamic collapse of the gas is not 
inhibited by the centrifugal barrier as 
in typical systems (e.g., typical 
disk galaxies are larger than their Schwarzschild radius
by 6-8 orders of magnitude because of rotational support).  
In Paper II we show that the existence of low-spin systems at high redshifts 
can in principle account for the seeds of quasar black holes
with masses $\gsim 10^6 M_\odot$ and a comoving density of bright galaxies. 
Appendix A then shows that
if a black hole forms in the initial collapse of a high-$\sigma$ peak on the
$\sim10^{8}\,\rm M_\odot$ mass scale, 
it is likely to be surrounded by a 
highly overdense region even on the mass scale of a galactic bulge. 
Due to the proximity of the centers of mass of the two systems, the black hole
will sink by dynamical friction to the center of the bulge system after it forms.
The later 
collapse of the surrounding region would fuel the black hole and result in the quasar activity
(Loeb \& Rasio 1994).
In the application of the ellipsoid model to
the formation of quasar black holes, we are free to consider only the inner
part of the collapse and to neglect subsequent accretion, since the formation
of the seed black hole occurs prior to this accretion.  
The main remaining uncertainty in this treatment is the omission of 
higher multipole torque couplings.

Given a primordial power-spectrum of Gaussian density perturbations,
the ellipsoid model makes definite predictions about the 
statistics of shapes and angular momenta of overdense regions in the universe. 
It would first be useful to compare the quantitative predictions
of our model to the statistical properties of collapsing regions
in high-resolution simulations (Bertschinger 1993 and references therein). 
Such a comparison would require large simulated volumes 
($\gsim [100 {\rm Mpc}]^3$)
in order to obtain reasonable 
statistical samples of rare objects that collapse today ($M\gsim 10^{15} M_\odot$).
The prominence of filaments and 
sheets relative to quasi-spherical structures
could also be tested observationally using
galaxy surveys (e.g., Geller \& Huchra 1989; 
Maddox et al. 1990; Saunders
et al. 1991; Shectman et al. 1992; Strauss et al. 1992; or the future DSS survey,
Gunn \& Knapp 1993)
to infer the smoothed mass distribution on scales larger than the
virialized cores of clusters of galaxies.  A non-spherical analysis of this type
can provide constraints which are complementary to the 
conventional results obtained by applying spherical filters to 
galaxy surveys.

\bigskip
\bigskip
\noindent
We thank
John Dubinski and David Weinberg for useful discussions.
D.J.E. was supported in part by a
National Science Foundation Graduate Research Fellowship.

\np

\noindent

\centerline{\bf APPENDIX A: PROBABLILITY DISTRIBUTION FOR THE INITIAL DATA}

Our model has $6N+5$ real numbers as its initial data, each of which is defined
as some integral over the initial density field.  
In this Appendix we derive the probability distribution
for these initial data in terms of the properties of the Gaussian random field
of initial density perturbations.

We denote the radii of the boundaries between the shells as 
$R_0, R_1, \ldots, R_N$, where $R_N=\infty$.  The values to 
be derived from the field are ($m=0,\pm1,\pm2$):
$$
\overline\delta(R_n) = \left({3\over4\pi R_n^3}\right) 
\int_{|{\bf r}|<R_n}{d^3r\,\delta({\bf r})}{{\rm \quad for}\;n=0,1,\ldots,N-1;}
\eqno\rm(A1)
$$
$$
q_{2m} = \rho_b\int_{|{\bf r}|<R_0}{d^3r\,
\delta({\bf r})\,r^2Y_{2m}^*};
\eqno\rm(A2)
$$
$$
b^{(n)}_{2m} = \rho_b \int_{R_n<|{\bf s}|<\infty}{d^3s\, Y^*_{2m} 
\delta({\bf s}) s^{-3}}{{\rm \quad for}\;n=0,1,\ldots,N-1.}
\eqno\rm(A3)
$$
>From these values, we can easily transform back to our model input variables in
equation $\labp$ and $\labn$, noting
$$a^{(n)}_{2m} = b^{(n-1)}_{2m} - b^{(n)}_{2m}{{\rm \quad for}\;n=1,2,\ldots,N-1.}
\eqno\rm(A4)
$$

Let us first switch to a multipole expansion of the Gaussian 
random field, as described by Binney \& Quinn (1991).  We replace the random field
$\delta({\bf x})$ by a new set of random functions $\{\delta_{\ell m}(k)\}$:
$$
\delta({\bf x}) = \sqrt{2\over \pi}\sum^\infty_{\ell=0}\sum^\ell_{m = -\ell}
\int^\infty_0{dk\,\delta_{\ell m}(k)\,kj_\ell(kr)Y_{\ell m}(\theta,\phi)},
\eqno\rm(A5)
$$
where $j_\ell$ are the spherical Bessel functions and $(r,\theta,\phi)$ 
are spherical coordinates.
Binney \& Quinn (1991) show that the set of functions  $\{\delta_{\ell m}\}$
is Gaussian distributed as
$$P[\{\delta_{\ell m}(k)\}] \propto \prod_{\ell, m} \exp\left[-\int^\infty_0{dk\, 
{|\delta_{\ell m}(k)|^2\over 2P(k)}}\right],\eqno\rm(A6)$$
where $P(k)$ is the power spectrum.
This means that the functions $\delta_{\ell m}(k)$ are independent and that all
of them have the same simple probability distribution.

When we insert equation (A5) into equations (A1)-(A3), the angular integrals
may be easily done.  The $q_{2m}$ and $b^{(n)}_{2m}$ coefficients depend only on
$\delta_{2m}(k)$, and $\overline\delta(R_n)$ depends only on
$\delta_{00}(k)$.  All the other $\ell$ in the expansion do not appear, so we
do not need to determine those $\delta_{\ell m}(k)$.  Furthermore,
the values we seek fall into six independent sets, since none of the integrals in 
equations (A1)-(A3) mix different $(\ell,m)$ sets.  

We next perform the radial integrals by using the identities 
(Abramowitz \& Stegun 1957)
$$
\eqalign{
{d\over dz}\left[z^{n+1}j_n(z)\right] &= z^{n+1} j_{n-1}(z) ,\cr
{d\over dz}\left[z^{-n} j_n(z)\right] &= -z^{-n} j_{n+1}(z).\cr}\eqno\rm(A7)
$$
We thus find
$$
\eqalign{\overline\delta(R_n) &= {3\over4\pi R_n^3}
\sqrt{2\over \pi} \int^\infty_0{dk
\int^{R_n}_0{dr\,\delta_{00}(k)\,r^2 k j_0(kr)}}\cr
&= {3\over4\pi R_n^3} \sqrt{2\over \pi} \int^\infty_0{dk\,k^{-2}
\delta_{00}(k) \int^{kR_0}_0{dz \,z^2 j_0(z)}}\cr
&= {3\over4\pi R_n^3} \sqrt{2\over \pi} \int^\infty_0{dk\,k^{-2}
\delta_{00}(k) (kR_n)^2 j_1(kR_n)} \cr
&= {3\over \pi\sqrt{2}} \int_0^\infty{dk\,\delta_{00}(k) {j_1(kR_n)\over R_n}}. \cr}
\eqno\rm(A8)
$$
Similarly, 
$$
q_{2m} = \rho_b R_0^5\sqrt{2\over \pi} \int_0^\infty {dk\,\delta_{2m}(k)
{j_3(kR_0)\over R_0} }\eqno\rm(A9)$$
and
$$
b^{(n)}_{2m} = \rho_b \sqrt{2\over \pi} \int_0^\infty {dk\,\delta_{2m}(k)
{j_1(kR_n)\over R_n} } .
\eqno\rm(A10)
$$
Each desired quantity has been written as an integral over the $\delta_{\ell m}(k)$,
a property that will be used later in finding the relevant probability distribution.

Before developing the joint probability distribution for the full problem, 
we wish to illustrate the method by working a simpler example first.
We consider the average overdensities inside two different
radii and ask: given a value for the overdensity inside the smaller
radius, what is the distribution for that at the larger radius?  Denoting
the radii $R_1$ and $R_2$, we have from equation (A9):
$$
\eqalign{\overline\delta_1 &= {3\over \pi\sqrt{2}} \int_0^\infty{dk\,\delta_{00}(k) 
{j_1(kR_1)\over R_1}}  ,\cr
\overline\delta_2 &= {3\over \pi\sqrt{2}} \int_0^\infty{dk\,\delta_{00}(k) 
{j_1(kR_2)\over R_2}}.\cr}
\eqno\rm(A11)
$$

Because the functions $\delta_{\ell m}(k)$ are independent 
and because we need only consider one at a time, we will suppress the $\ell m$ label.
The form of the integral in equation (A6) prompts us to consider $\delta(k)$ 
as a vector in a function space
and to define the inner product as
$$\left<a|b\right> = \int^\infty_0{{dk\over P(k)}\,a^*(k)b(k)},\eqno\rm(A12)$$
for two arbitrary functions $a$ and $b$.
Next, define:
$$\eqalign{Q_1(k) &= {j_1(kR_1)\over R_1}P(k)  ,\cr
Q_2(k) &= {j_1(kR_2)\over R_2}P(k).\cr}\eqno\rm(A13)$$
Then we have 
$$\overline\delta_1 = {3\over \pi\sqrt2}\left<Q_1|\delta_{00}\right>\eqno\rm(A14)$$
and similarly for $\overline\delta_2$; note that the $P(k)$ in the definition 
of $Q$ cancels the $P(k)$ in the inner product.

Now let us consider some basis $n_j(k)$ that is orthonormal under our inner 
product.  Then we may write $\delta(k)$ as a linear combination of these basis
vectors,
$$\delta(k) = \sum_j{\beta_j n_j(k)}.\eqno\rm(A15)$$
Substituting this form into equation (A6), we find the probability for a given
set of $\beta_j$ to occur,
$$P({\beta}) \propto \exp\left(
-{1\over2}\sum_{i,j}\beta_i^*\beta_j\left<n_i|n_j\right>\right) =
\exp\left(-{1\over2}\sum_j |\beta_j|^2\right)=
\prod_j\exp\left(-{1\over2}|\beta_j|^2\right).\eqno\rm(A16)$$
This means that each $\beta$ is independently distributed with a uniformly distributed
phase and a Gaussian distributed magnitude.  For an $m=0$
function as we have here, $\beta$ is real and is simply a Gaussian deviate, 
namely it is drawn from a Gaussian
distribution with a zero mean and a unit variance.

We would like to express $Q_1$ and $Q_2$ in terms of this orthonormal basis.  
Since we have not yet constructed the basis, we choose it to be simple.
The first vector $n_1$ will be $Q_1$ normalized, 
$$
n_1 = {Q_1\over \sqrt{\left<Q_1|Q_1\right>}}.
\eqno\rm(A17)
$$
Then, we pick $n_2$ to be the normalized component of $Q_2$ orthogonal to $n_1$, 
which is
$$
n_2 = {Q_2 - \left<n_1|Q_2\right>n_1 \over
\sqrt{\left<Q_2|Q_2\right>-\left<n_1|Q_2\right>^2} }.
\eqno\rm(A18)
$$
We may then complete the basis any way we wish; none of the other vectors will
include any component of $Q_1$ or $Q_2$.
Inverting (A17) and (A18), we find
$$
\eqalign{Q_1 &= \sqrt{\left<Q_1|Q_1\right>}n_1,
\cr 
Q_2 &= \left<n_1|Q_2\right>\,n_1 + 
\sqrt{\left<Q_2|Q_2\right>-\left<n_1|Q_2\right>^2}\, n_2.\cr}
\eqno\rm(A19)
$$

Putting equations (A9) and (A15) into (A14) and using the orthonormality
of the basis, we find
$$\overline\delta_1 = {3\over \pi\sqrt2} \sqrt{\left<Q_1|Q_1\right>}\,\beta_1
\eqno\rm(A20)$$
and
$$\eqalign{\overline\delta_2 &= {3\over \pi\sqrt2} 
\left(\left<n_1|Q_2\right>\,\beta_1 + 
\sqrt{\left<Q_2|Q_2\right>-\left<n_1|Q_2\right>^2}\, \beta_2\right)\cr
&= {3\over \pi\sqrt2}\sqrt{\left<Q_2|Q_2\right>}\left( \gamma\beta_1 +
\sqrt{1-\gamma^2}\beta_2\right),\cr}\eqno\rm(A21)$$
where $\gamma$ is the dimensionless overlap defined as
$$\gamma = {\left<Q_1|Q_2\right>^2\over\sqrt{
\left<Q_1|Q_1\right>\left<Q_2|Q_2\right>}}.\eqno\rm(A22)$$
To determine $\overline\delta_1$ and $\overline\delta_2$ randomly, we simply
find two Gaussian deviates $\beta_1$ and $\beta_2$ and combine them as indicated.
Alternatively, if we are given $\overline\delta_1$ and asked to determine 
$\overline\delta_2$ given this constraint, we fix the value of $\beta_1$
to produce $\overline\delta_1$.  Despite this constraint, $\beta_2$ remains
a Gaussian deviate since it is independent of $\beta_1$, and so we may
take a random value of $\beta_2$ along with our fixed value of $\beta_1$ to
find the constrained distribution of $\overline\delta_2$.

Equation (A20) can be expanded to give
$$
(\overline\delta_1)^2 = {1\over 2\pi^2}\int_0^\infty{dk\,P(k) 
\left({3j_1[kR_1]\over R_1}\right)^2} \beta_1^2,\eqno\rm(A23)
$$
the expectation value of which is the standard expression for $\left<(\delta M/M)^2\right>$
for a spherical top-hat window function of radius $R_1$ [cf. Eq. $\labv$].
Denoting this expectation value as $(\sigma_1)^2$ and similarly for $(\sigma_2)^2$, we
find $\nu_1\equiv\overline\delta_1/\sigma_1 = \beta_1$ and $\nu_2 \equiv
\overline\delta_2/\sigma_2 = \gamma\beta_1 + \sqrt{1-\gamma^2}\beta_2$, so
a high-$\sigma$ peak on one mass scale will correspond to a high-$\sigma$ peak
on the other mass scale if $\gamma$ is close to 1.  

For the $\Omega=1$ CDM model and
mass scales of $10^8\,M_\odot$ and $10^{10}\,M_\odot$, we calculate 
$\gamma\approx0.74$.  
This means, for example, that given a $3\sigma$ peak at 
$10^8\,\rm M_\odot$, the surrounding $10^{10}\,\rm M_\odot$ region has
an overdensity of $\nu_2 = 2.22 \pm 0.67\beta$, where $\beta$ is a Gaussian deviate.
This result has interesting physical implications to the problem
of the origin of quasar black holes discussed in Paper II.
If a black hole forms in the initial collapse of a high-$\sigma$ peak on the
$\sim10^{8}\,\rm M_\odot$ mass scale, 
it is likely to be surrounded by a 
highly overdense region even on the mass scale of a galactic bulge. 

We now need to generalize the above method to an arbitrary number of vectors.
We seek $p$ values from the field, denoted $f_1,f_2,\ldots,f_p$, and each
may be written as some integral over a particular $\delta_{\ell m}(k)$;
only one $(\ell,m)$ is involved and so we suppress the index here as well.
Then, we define the $Q_j(k)$ by the relations
$$
f_j = \int_0^\infty{{dk\over P(k)}\,\delta(k)Q_j(k)},
\eqno\rm(A24)
$$
where all of the $Q_j$ are real-valued functions;
and define the matrix $M$,
$$
M_{ij} = \left<Q_i|Q_j\right> = \int_0^\infty{{dk\over P(k)}\, Q_i(k)Q_j(k)}.
\eqno\rm(A25)
$$
We now seek an orthonormal basis $\{n_j\}$ with the property that 
$$
Q_i = \sum^p_{j=1}L_{ij}n_j,
\eqno\rm(A26)
$$
where $L$ is a lower triangular matrix and the remaining $n_j$ are not used.  
Substituting equation (A26) into (A25) and using the orthonormality 
property, we find
$$
M_{ij} = \sum_p{L_{ip}L_{jp}}.\eqno\rm(A27)
$$
Hence, our basis must be defined by $M = LL^T$.  Can we find such an $L$?
In fact we can, for if $M$ is postive-definite symmetric matrix, 
this is the Cholesky decomposition (Press \etal 1992, \S 2.9).  
$M$ is obviously symmetric and must be positive-definite if $P(k)>0$ for
all $k$, and so we may find a unique $L$.

Next, we take $\{n_j\}$ as the basis for $\delta(k)$.  Then, as before, we 
decompose by equation (A15) and reach the probability distribution in 
(A16).  Substituting this result, as well as equation (A26), into (A24), 
we get 
$$
f_j = \int_0^\infty{{dk\over P(k)}\,\sum_{p,q}\beta_p n_p L_{jq} n_q}
= \sum_p L_{jp} \beta_p.
\eqno\rm(A28)
$$
This completes the solution. One needs only to specify $p$ Gaussian 
deviates and combine them in the indicated way in order to get realizations of the
$p$ values ${f_j}$.  To construct the $L$ matrix, one must perform $p(p+1)/2$ one-dimensional
integrals as given in equation (A25). This task can be made even simpler by
doing the integrals together since they have many functional evaluations
in common.

In the particular case of the initial conditions for the ellipsoid model,
we have six sets of parameters to determine, one for $\ell=0$ and five
for $\ell=2$.  But the five $\ell=2$ sets are equivalent and will share
one $L$ matrix.  Also, the integral forms for $\ell=0$ in equation (A8)
are identical to those for $\ell=2$ as given in equation (A11).  The
only difference between the two sets is the one additional initial condition
presented in equation (A9).  In order to compute $L$ for the $\ell = 0$ set,
we need not calculate any new integrals; we simply take the required integrals
as a submatrix of $M$ from the $\ell=2$ calculation, and compute a new 
Cholesky decomposition.

Because the original field $\delta({\bf x})$ was real, the 
complex functions $\delta_{\ell m}(k)$ have the usual spherical
harmonic relation between them, i.e. 
$\delta_{\ell m}(k) = (-1)^m\delta^*_{\ell\,-m}(k)$.
Thus, for $m\neq0$ there are four real functions:
the real and imaginary parts of $\delta_{\ell m}(k)$
and $\delta_{\ell -m}(k)$, only two of which are 
independent.  Returning to equations $\labr$ and $\labq$, where $\ell=2$, we choose
to work with the real and imaginary parts of $m=1$ and $m=2$ as well as the 
real term $m=0$ as the five independent real parameters.  This means
that instead of asking for $q_{2m}$ and $b^{(n)}_{2m}$ as in equations (A2) and
(A3), we compute ${\rm Re\,}q_{22}$, ${\rm Im\,}q_{22}$, 
${\rm Re\,}q_{21}$, ${\rm Im\,}q_{21}$, and $q_{20}$, and the same for $b^{(n)}$.
This is easy to do by simply taking the 
quantities ${\rm Re\,}\delta_{22}(k)$, ${\rm Im\,}\delta_{22}$, ...,  
as the independent random functions.  The $\ell=2$ values again fall into five 
independent sets, each being treated as above.  There is however one modification, as 
seen from the original multipole field probability distribution in equation (A6):
$$
\eqalign{P[\delta({\bf x})] &\propto \exp\left[ -\sum_{\ell,m}
\int_0^\infty{dk\,{|\delta_{\ell m}|^2\over 2P(k)}}\right]\cr
&= \exp\left[ -\sum_{\ell} \int_0^\infty{dk\,{2({\rm Re\,}\delta_{\ell\ell})^2 + 
2({\rm Im\,}\delta_{\ell\ell})^2 + 2({\rm Re\,}\delta_{\ell(\ell-1)})^2 + \ldots +
(\delta_{\ell 0})^2 \over 2P(k)}}\right].\cr}
\eqno\rm(A29)
$$
This indicates that the real and imaginary parts of the $m\ne0$ terms actually
are distributed with half the variance of the $m=0$ term.  The simplest way
to incorporate this fact is to compute all the $\ell=2$ sets as for the $m=0$ case and
then take the $\beta_j$ to have variance of one-half rather than 
unity for $m\ne0$.

Finally, there is a choice as to what order to place our $f_j$
in the vector.  Since the matrix $L$ is a lower triangular, the value $f_1$ requires
only one Gaussian deviate to be determined.  As discussed in \S 4, 
there are additional constraints that we place on the initial conditions.
Since these are imposed simply by rejecting any set of initial data that do not
satisfy them, it is most efficient to pick the most discriminating test to be $f_1$.
In particular, the test that $\delta(R_0)>\nu_{min}\sigma$ is very stringent
and we arrange the $\ell=0$ functions so that this condition is the first
basis function.  This way, we need only find one Gaussian deviate before
applying this test. 

\np
\centerline{\bf APPENDIX B: TREATMENT OF AN AXIS COLLAPSE}

The inclusion of a collapsed axis
is complicated by the facts that the ellipsoid is rotating and that the 
columns of the matrix $A$ need not be orthogonal.  When
we halt the collapse of an axis, we wish only to end the radial collapse and to
leave the tangential velocity unchanged.

First, we must identify that an axis has reached 40\% of its turn-around value.
At each time step, we diagonalize $AA^T = Q\Lambda Q^T$ 
to find the axis lengths, which are in $\sqrt\Lambda$.
We put these in ascending order and compare each to the previous maximum length
for the short, middle, and long axes.  If the new length is longer, we save it
as the new maximum.  If the new length is less than 40\% of the maximum for that
axis, we assume that the axis has collapsed and save the column number so as to be
able to refer to it in $Q$ and $\Lambda$.  This method of identifying
the axes by sorting the lengths works well because the axes are well-separated
soon after the beginning of the integration and well before turn-around.

If a particular axis has collapsed, we remove the radial component of 
the velocity in that direction and we alter the right hand side of the
equation of motion $\labg$ so that there is no inward force in that
direction.  The steps for each alteration are equivalent, so we only describe the first in
detail.

Suppose we begin from the diagonalization $AA^T = Q\Lambda Q^T$ and want to fix the $\alpha$-th
axis.  The position of mass elements within the ellipsoid may be written as
${\bf r} = Q\sqrt\Lambda{\bf s}$, where ${\bf s}$ is a vector on the unit sphere.
Then the frozen axis is ${\bf s} \parallel {\bf e}_\alpha$, the Cartesian basis vector.
The velocity field is ${\bf v} = \dot{A}A^{-1}Q\sqrt\Lambda{\bf s}$, where
$\dot A = dA/dt$.  We now rotate this velocity to the principal axis 
frame of the ellipsoid, which requires left-multiplying by $Q^T$.  This is obtained by
defining 
$$
\tilde V = Q^T\dot{A} A^{-1}Q\sqrt\Lambda  ,
\eqno\rm(B1)
$$
where $\tilde V_{jk}$ is the $j$th component of the velocity at 
the point in the direction of the $k$th axis, and the axes are numbered
according to the columns of $Q$.

Now we can remove the $\alpha$-th component of velocity along the $\alpha$-th 
axis by removing the $\alpha\alpha$ component of $\tilde{V}$.  We therefore define a new matrix
$W$ such that
$$
U_{\beta\gamma} = \cases{0 &if $\beta=\alpha$ and $\gamma=\alpha$,\cr 
\tilde V_{\beta\gamma} &otherwise.\cr}\eqno\rm(B2)$$
Then we need to connect this matrix back to the new velocity matrix $\dot A_{new}$.
We do this by solving equation (B1) for $\dot A$ after replacing $\tilde V$ by
$U$.  Then we replace the old matrix for $\dot A$ with
$$
\dot A_{new} = QU{\sqrt\Lambda}^{-1}Q^TA.\eqno\rm(B3)
$$
If more than one axis has collapsed, we
modify the definition of $U$ to eliminate all diagonal elements corresponding to the
collapsed axes.
This algorithm works
equally well for the forces by 
substituting  $\ddot A\equiv d^2A/dt^2$ for $\dot A$
everywhere.  Since this procedure does not modify the tangential velocities,
it preserves angular momentum; however it does change the vorticity.

\bigskip
\bigskip
\centerline{\bf APPENDIX C: RELATION BETWEEN SHEAR AND OVERDENSITY}

In this Appendix we present a calculation separate from the dynamical
ellipsoid model, but yet based upon the formalism discussed
in this paper.  As before, we begin by dividing the universe into two pieces
through a spherical boundary of radius $R$ around the origin.  We now wish
to compare the statistical properties of the average overdensity in the 
interior region $\overline\delta(R)$ [cf. Eq. $\labo$] to those
of the quadrupole shear resulting from the exterior region $a_{2m}$ 
[cf. Eq. $\labf$].  We will show that these quantities are drawn from 
independent Gaussian distributions whose variances are related by
a constant of proportionality that is independent of the power spectrum or
mass scale.

>From equations (A1), (A3), (A8), and (A11), we 
can easily relate $\overline\delta$ and $a_{2m}$ to integrals over the 
coefficients of the multipole expansion of the Gaussian random field:
$$
\overline\delta = {3\over\pi\sqrt2} \int_0^\infty{dk\,\delta_{00}(k){j_1(kR)\over R}}
\eqno\rm(C1)$$
and
$$
a_{2m} = \rho_b\sqrt{2\over\pi}\int_0^\infty{dk\,\delta_{2m}(k){j_1(kR)\over R}}.
\eqno\rm(C2)$$
Because each of these six quantities depends upon different $\delta_{\ell m}$,
they will be independent.  By the methods of Appendix A, 
we find that the six quantities are
Gaussian distributed with a zero mean and the variances
$$
\left<\overline\delta^2\right> = \left(3\over\pi\sqrt2\right)^2\int{dk\,P(k) 
\left(j_1(kR)\over R\right)^2},
\eqno\rm(C3)
$$
$$
\left<(a_{20})^2\right> = \left(\rho_b\sqrt{2\over\pi}\right)^2\int{dk\,P(k) 
\left(j_1(kR)\over R\right)^2},
\eqno\rm(C4)
$$
and
$$
\left<({\rm Re\,}a_{22})^2\right> = \left<({\rm Im\,}a_{22})^2\right> = 
\left<({\rm Re\,}a_{21})^2\right> = \left<({\rm Im\,}a_{21})^2\right> = 
{1\over 2} \left<(a_{20})^2\right>.
\eqno\rm(C5)
$$
The last equation follows from equation (A29) and the discussion 
around it.

The key result of this derivation is that the integrals in equations (C3) and
(C4) are identical.  We can therefore pull all of the dependence on the
power spectrum into one constant.  We define $\sigma$ as the {\it r.m.s.} amplitude of
$\delta M/M$ according to equation $\labv$.
We may then write 
$\overline\delta = \sigma\nu$, $a_{20} = (2\sqrt\pi\rho_b/3)\sigma z_1$,
${\rm Re\,}a_{22} = (\sqrt{2\pi}\rho_b/3)\sigma z_2$, 
${\rm Im\,}a_{22} = (\sqrt{2\pi}\rho_b/3)\sigma z_3$, 
${\rm Re\,}a_{21} = (\sqrt{2\pi}\rho_b/3)\sigma z_4$, and
${\rm Im\,}a_{21} = (\sqrt{2\pi}\rho_b/3)\sigma z_5$, 
where $\nu$ and the $\{z_j\}$ are all Gaussian deviates of unit variance.

We next consider how the inner region would evolve given these values as
the initial conditions.  In particular, we apply the Zel'dovich approximation
to find whether all the material of the interior region will 
turn-around and form a spatially bounded object or whether the region will
expand forever in at least one direction.  For an $\Omega=1$ universe and under
the Zel'dovich approximation, this question is simply a matter of whether the initial
peculiar velocities have inward radial components or not.  This, in turn, depends
on the initial gravitational potential (as we assume growing mode velocities),
which we may construct from equations $\labs$ and $\labr$ to get the matrix
$$
\Phi_{\alpha\beta} =
{4\pi G\rho_b\sigma\over 3}(\nu I_{\alpha\beta} + S_{\alpha\beta}),\eqno\rm(C6)
$$
where $I$ is the identity matrix and 
$$
S = {\sqrt{3\over 5}}\pmatrix{z_2 - {z_1\over {\sqrt{3}}} & z_3 & z_4\cr
z_3 & -z_2 - {z_1\over {\sqrt{3}}} & z_5\cr
z_4 & z_5 & {2z_1\over {\sqrt{3}}}\cr}.\eqno\rm(C7)
$$
We want $\Phi$ to have positive eigenvalues, which means that the eigenvalues
of $S$ must be greater than $-\nu$.
The probability that a spherical region of overdensity $\nu$ will expand
without bound in one direction due to shear depends only on
the distribution of the random matrix $S$.  All reference to the power spectrum
has been removed.

Table 1 shows the cumulative distribution function for the most negative
eigenvalue $-\mu$ ($\mu>0$) of the matrix $S$.  
This quantity acts as a negative
$\nu$, counteracting the actual $\nu = \overline\delta/\sigma$ that one normally
identifies as the parameter controlling gravitational collapse. 
The importance of shear
through the collapse is evident from the prevalence of large values of $\mu$. 
For low $\Omega$ universes, one might
alter the peculiar velocity requirement as discussed in \S 4 
[cf. Eq. $\labw$]; this 
leads to the requirement that $\nu-\mu> \overline\delta_{crit}/\sigma$.

The above calculation provides insight to the acceptance rate for the velocity constraint
that we use in the ellipsoid model.  In this model, there is an additional
term in the potential coming from the anisotropy of the ellipsoid.  This term
tends to slightly increase the acceptance (i.e. resist the shear) 
because the long axis of the ellipsoid tends to be aligned with the 
most negative shear axis.  The above calculation also has implications to
the Press-Schechter formalism (Press \& Schechter 1974), in which spherical regions
with a sufficiently high $\nu$ are all assumed to turn into objects at the appropriate
collapse time.  However, as it stands the model can only eliminate objects on the
basis of shear, some of which would presumably be re-introduced through mergers
of lower mass systems or fragmentation of unbound filaments.

\np
\centerline{\bf REFERENCES}

\def\apj{ApJ}
\def\mnras{MNRAS}
\def\aap{{A\&A}}

\def\myref{\smallskip \noindent}

\myref
Abell, G. O., 1958, ApJ Suppl., 8, 211
\myref
Bardeen, J. M., Bond, J. R., Kaiser, N., \& Szalay, A. S.
1986, \apj, 304, 15
\myref
Barnes, J., \& Efstathiou, G. 1987, \apj, 319, 575
\myref
Bertschinger, E. 1993, IAS Preprint 93/68, submitted
to Physica D 
\myref
Bertschinger, E., \& Gelb, J. M. 1991, Comp. Phys., 5, 164
\myref
Bertschinger, E., \& Jain, B. 1994, \apj, in press
\myref
Binney, J., \& Quinn, T. 1991, \mnras, 249, 678
\myref
Binney, J., \& Tremaine, S. 1987, Galactic Dynamics (Princeton: Princeton Univ.
Press)
\myref
Birkinshaw M., \& Hughes J. P. 1994, ApJ, 420, 33 
\myref
Bond, J. R., \& Myers, S. T. 1993, \apj, submitted
\myref
Carlson, B. C. 1977, SIAM Journal on Mathematical Analysis, 8, 231
\myref
Cen, R., \& Ostriker, J. P. 1993, ApJ, 417, 415
\myref
Doroshkevich, A. G. 1970, Astrofizika, 6, 581
\myref
Dubinski, J. 1992, \apj, 401, 441
\myref
Eisenstein, D. J., \& Loeb, A. 1994, submitted to ApJ Letters (Paper II)
\myref
Efstathiou, G., \& Jones, B. J. T. 1979, MNRAS, 186, 133
\myref
Fukugita, M., Okamura, S., Yasuda, N. 1993, \apj, 412, L13
\myref
Geller, M. J., \& Huchra, J. P. 1989, Science, 246, 897
\myref
Gunn, J. E., \& Gott, J. R. 1972, \apj, 176, 1
\myref
Gunn, J. E., \& Knapp, G. R. 1993, in {Sky Surveys: Protostars to 
Protogalaxies}, ed. Soifer, B. T. (A.S.P. Conference Proceedings), 43, 267
\myref
Hoffman, Y. 1986, ApJ, 301, 65
\myref
Hoyle, F. 1949, in Problems of Cosmical Aerodynamics, ed. Burgers, J. M. 
\& van de Hulst, H. C. (Dayton: Central Air Documents Office), p. 195
\myref
Icke, V. 1973, \aap, 27, 1
\myref
Lin, C. C., Mestel, L., \& Shu, F. H. 1965, \apj, 142, 1431
\myref
Lynden-Bell, D. 1964, \apj, 139, 1195
\myref
Loeb, A., \& Rasio, F. A. 1993, \apj, in press
\myref
Maddox, S. J., Efstathiou, G., Sutherland, W. J., \& Loveday, J.
1990, MNRAS, 242, 43p 
\myref
McHardy, I. M., Stewart, G. C., Edge, A. C., Cooke, B. A., 
Yamashita, K., \& Hatsukade, I. 1990, MNRAS, 242, 215 
\myref
Merritt, D., \& Hernquist, L. 1991, ApJ, 376, 439
\myref
Park, C. 1990, MNRAS, 242, 59p
\myref
Peebles, P. J. E. 1969, \apj, 155, 393
\myref
Peebles, P. J. E. 1980, The Large Scale Structure of the Universe
(Princeton: Princeton Univ. Press)
\myref
Peebles, P. J. E. 1993, Principles of Physical Cosmology
(Princeton: Princeton Univ. Press)
\myref
Press, W. H. \& Schechter, P. 1974, \apj, 187, 425
\myref
Press, W. H., Teukolsky, S. A., Vetterling, W. T., \& Flannery, B. P. 1992,
Numerical Recipes in C, second edition (Cambridge: Cambridge Univ. Press)
\myref
Quinn, T., \& Binney, J. 1992, \mnras, 255, 729
\myref
Ryden, B. S. 1988, \apj, 329, 589
\myref
Saunders, W., Frenk, C.~S., Rowan-Robinson, M., Efstathiou, G.,
Lawrence, A.,  Kaiser, N., Ellis, R.~S., Crawford, J., Xia, X.-Y., \&
Parry, I. 1991, Nature, 349, 32
\myref
Shandarin, S. F., \& Zel'dovich, Ya. B. 1989, Rev. Mod. Phys., 61, 185 
\myref
Shectman, S. A., Schechter, P. L., Oemler, A. A., Tucker, D.,
Kirshner, R. P., \& Lin, H. 1992, in Clusters and Superclusters of
Galaxies, ed.~A.C. Fabian (Dordrecht: Kluwer)
\myref
Strauss, M. A., Davis, M., Yahil, A., \& Huchra, J. P. 1992, ApJ,
385, 421 
\myref
van de Weygaert, R., \& Babul, A. 1994, \apj, 425, L59
\myref
Warren, M. S., Quinn, P. J., Salmon, J. K.,
\& Zurek, W. H. 1992, \apj, 399, 405
\myref
White, S. D. M. 1984, \apj, 286, 38
\myref
White, S. D. M. 1993, private communication
\myref
White, S. D. M., \& Silk, J. 1979, \apj, 231, 1
\myref
Zel'dovich, Ya. B. 1965, Soviet Astron.-A.J., 8, 700 
\myref
Zel'dovich, Ya. B. 1970, \aap, 5 84
\np
\centerline{\bf FIGURE CAPTIONS}

\noindent
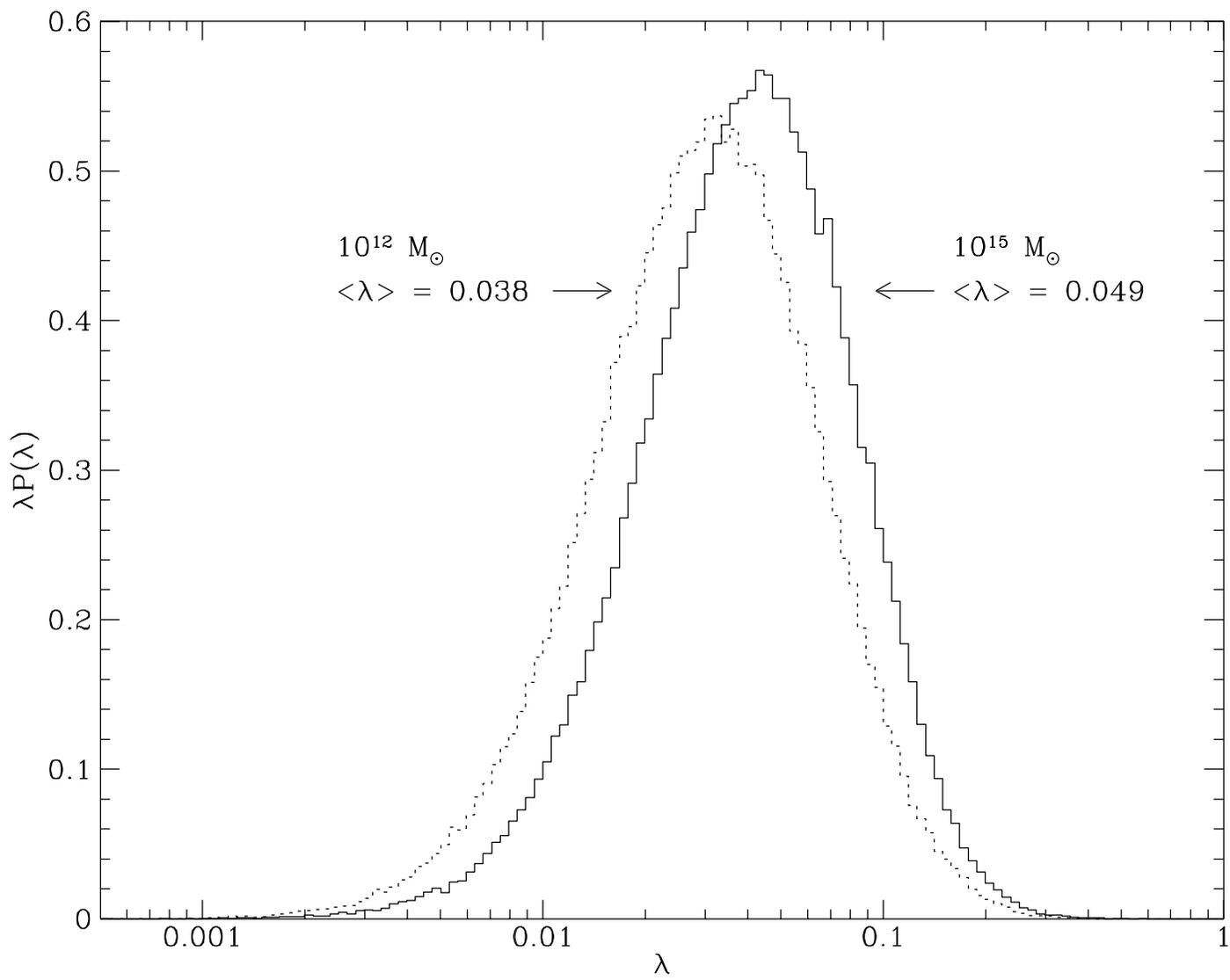
{\bf Fig. 1}: Probability distribution of the spin parameter $P(\lambda)$
for the mass scales of $10^{15}\,\rm M_\odot$
and $10^{12}\,\rm M_\odot$ and $\nu_{min} = 2.5$ in an $\Omega=1$ universe.
Each histogram includes $2\times10^5$ random realizations of the initial conditions. 
The average values $\langle\lambda\rangle=\int_0^\infty \lambda P(\lambda) d\lambda$
are also shown.

\noindent
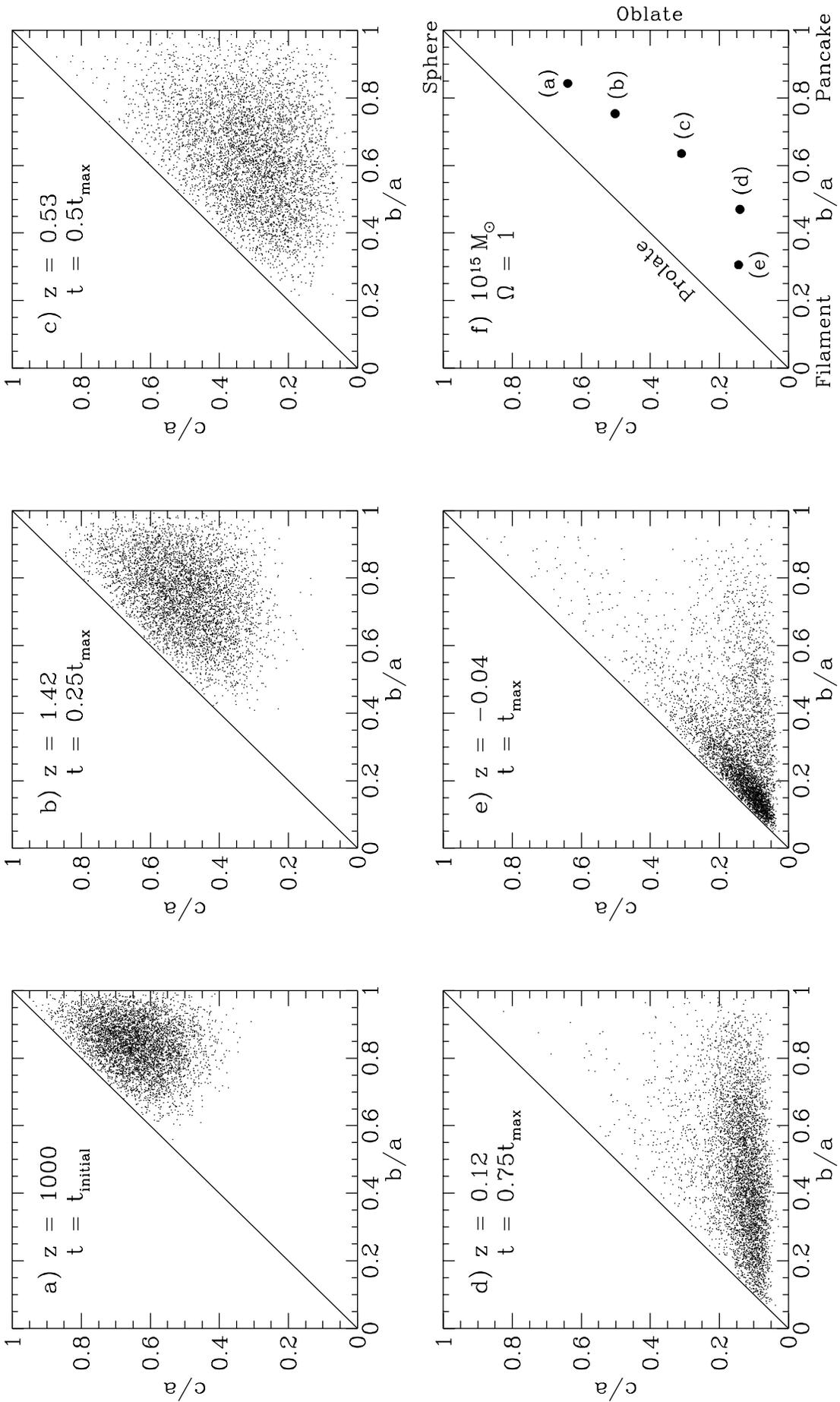
{\bf Fig. 2}: Distribution of axis ratios for an ensemble of 5000 systems
of mass $10^{15}\,\rm M_\odot$ and $\nu = 2$ in an $\Omega=1$ universe.
Panel (a) shows the initial distribution and panels (b)-(d) show the 
resulting evolution at particular times.  Panel (e) shows the
distribution at time $t_{\rm max}$, the end time of
the integration defined by the collapse time of a homogeneous spherical perturbation with
the same initial density as the ellipsoids.  Panel (f) shows the mean
values of the distributions in the other panels.  We define the axis lengths
as $a\geq b\geq c$.  To find the redshifts we assumed 
a bias parameter of $1.3$ and $H_0= 50\,\rm km/s/Mpc$.

\noindent
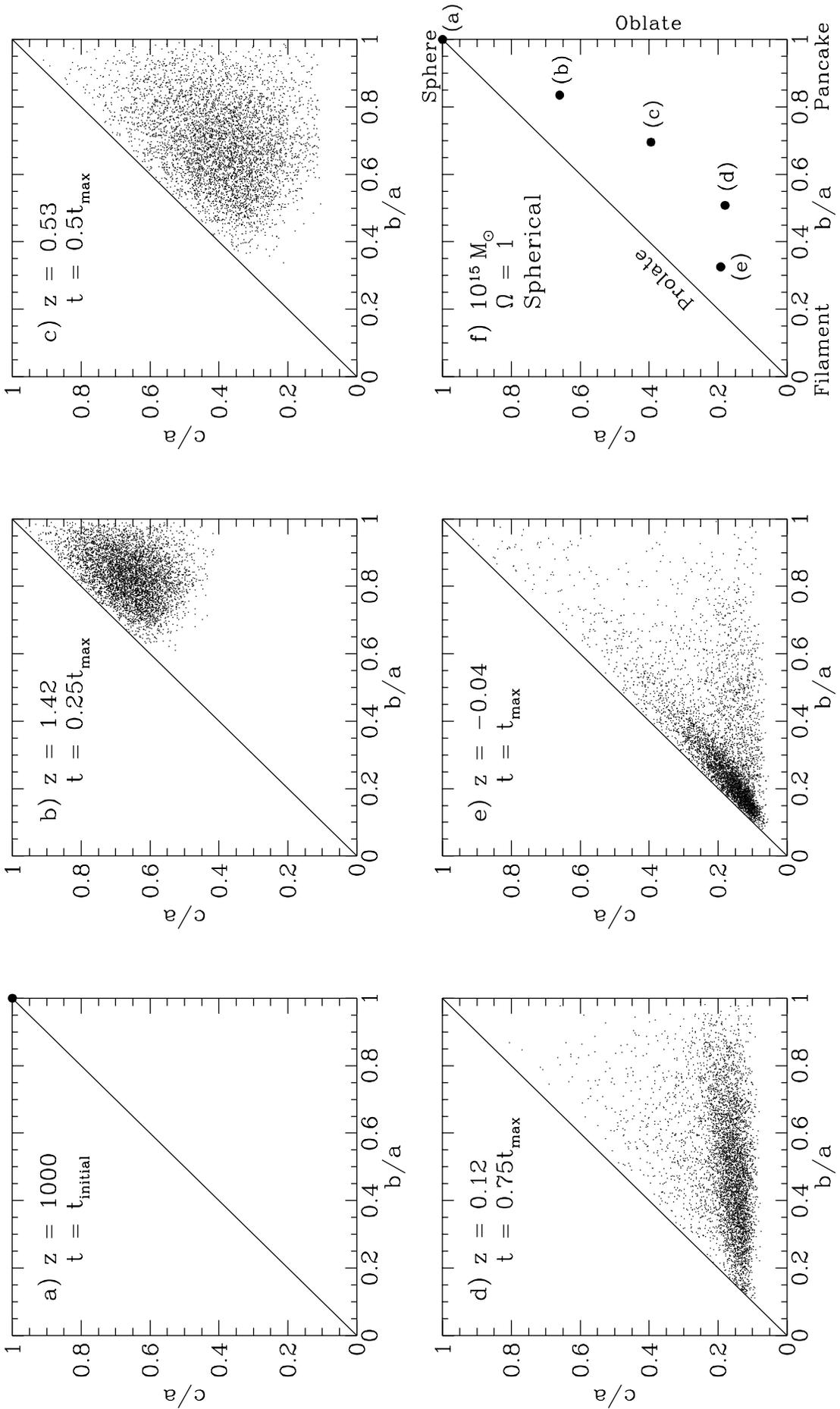
{\bf Fig. 3}: As in Figure 2, but with the initial ellipsoid replaced by a sphere
of the same density.  The external shear was unchanged.  

\noindent
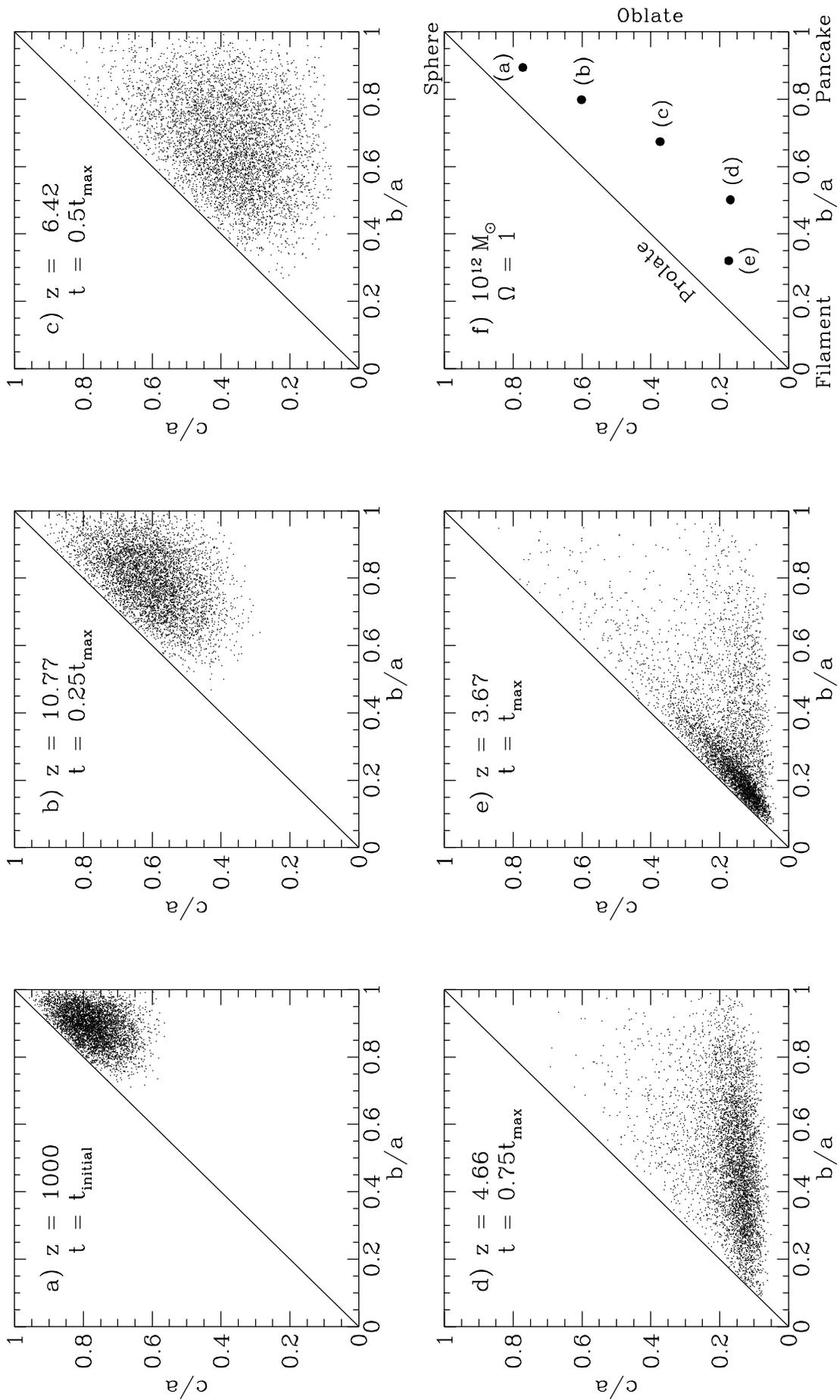
{\bf Fig. 4}: As in Figure 2, but for $10^{12}\,\rm M_\odot$ regions.  

\noindent
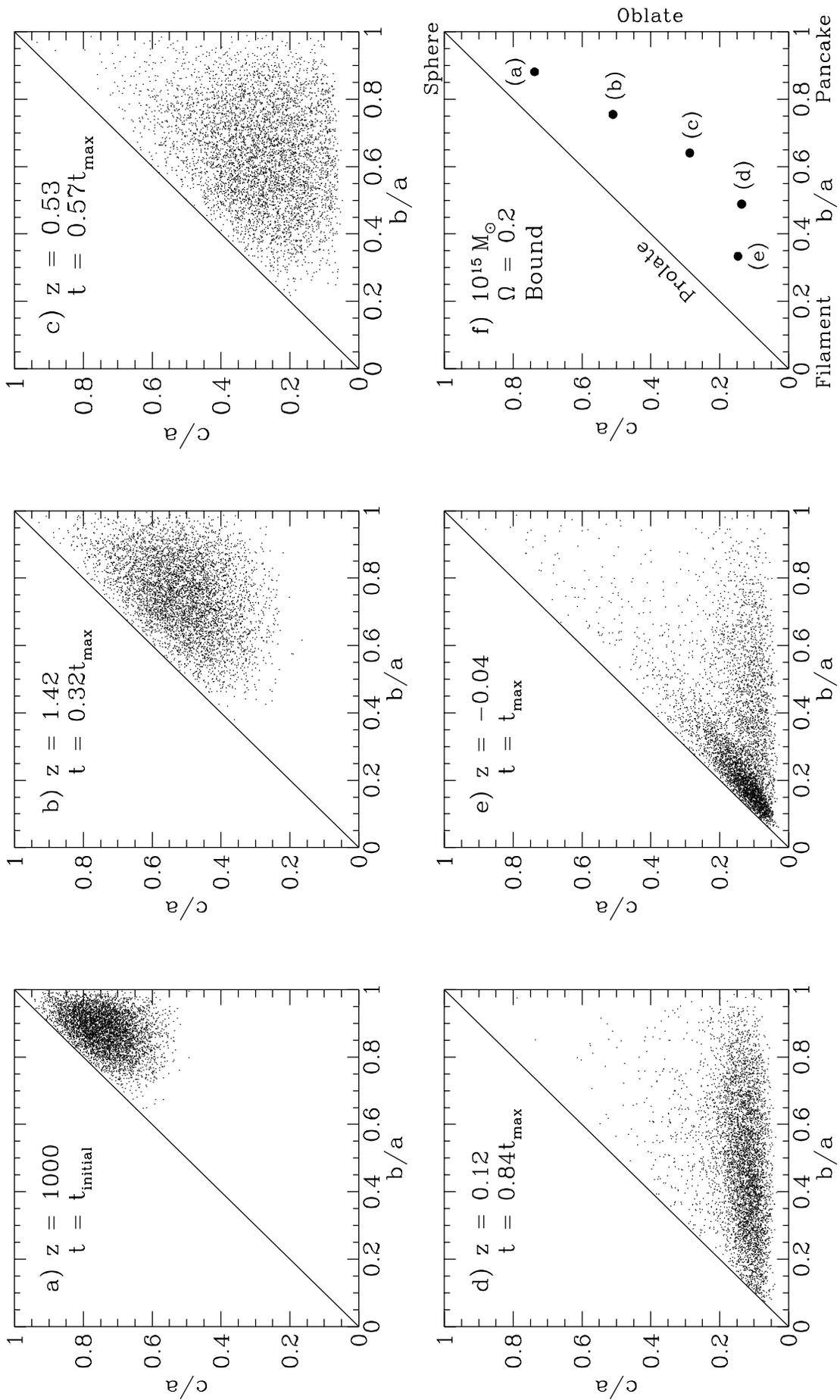
{\bf Fig. 5}: As in Figure 2, but for an open universe with $\Omega = 0.2$, 
$H_0 = 50\,\rm km/s/Mpc$, and no bias ($b=1.0$).  For $10^{15}\,\rm M_\odot$
ellipsoids, we pick $\nu = 2.2$ in order to match the final collapse redshift 
of Figure 2 and then pick the time slices to match the other redshifts of that
figure.

\noindent
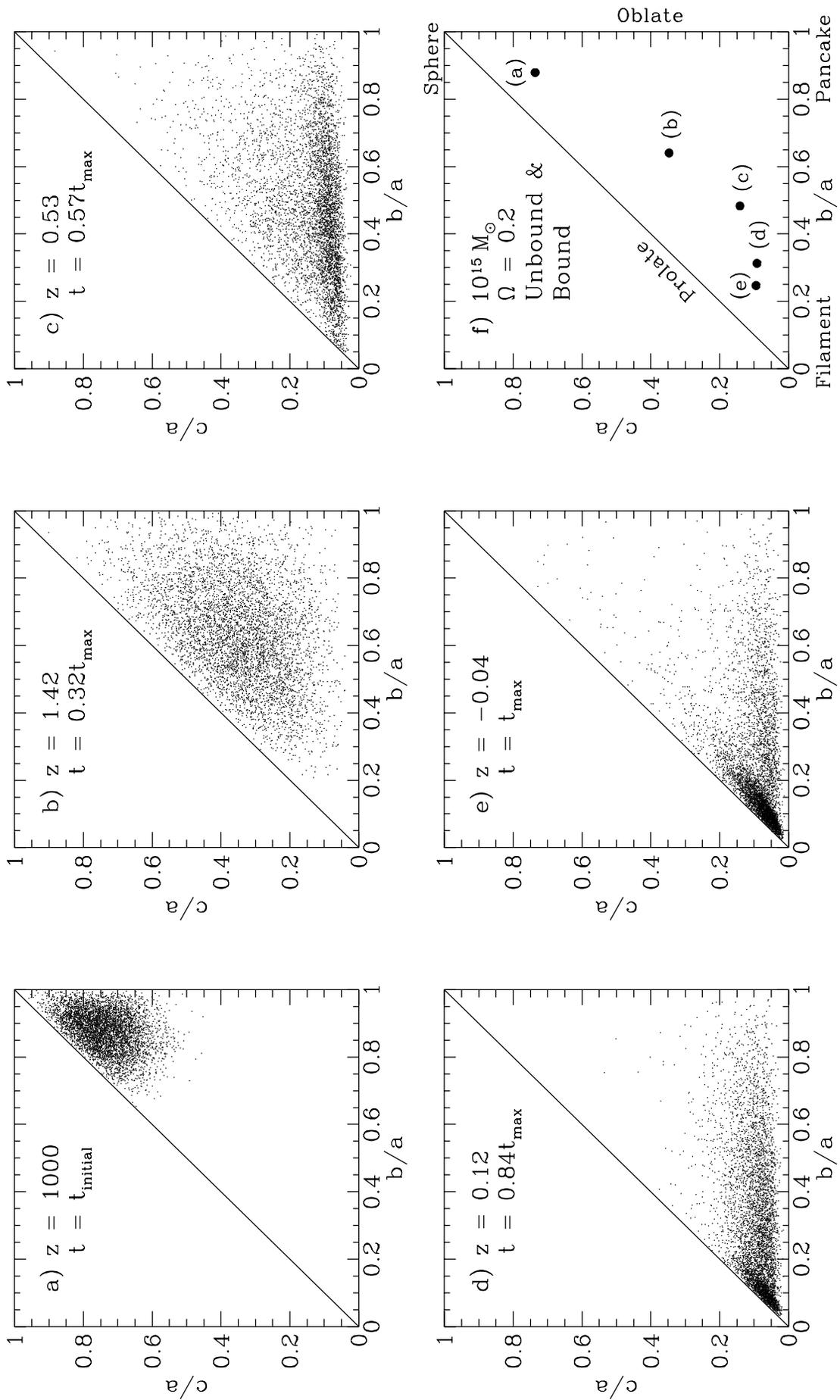
{\bf Fig. 6}: As in Figure 5, but after removing the constraint on the initial
peculiar velocities.  This allows the inclusion of objects with at least one
axis that is likely to expand without bound.

\noindent
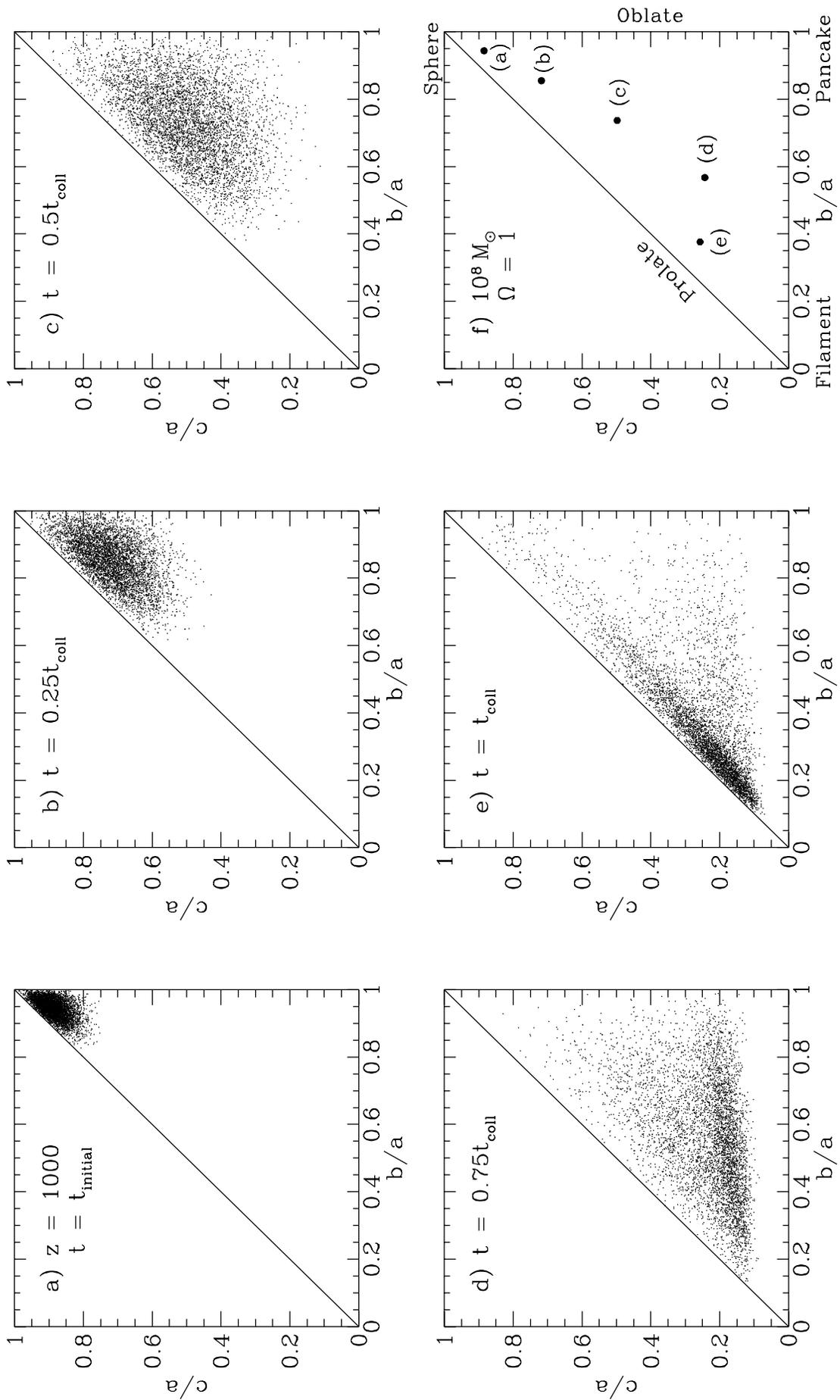
{\bf Fig. 7}: As in Figure 2, but for $10^{8}\,\rm M_\odot$ regions.

\noindent
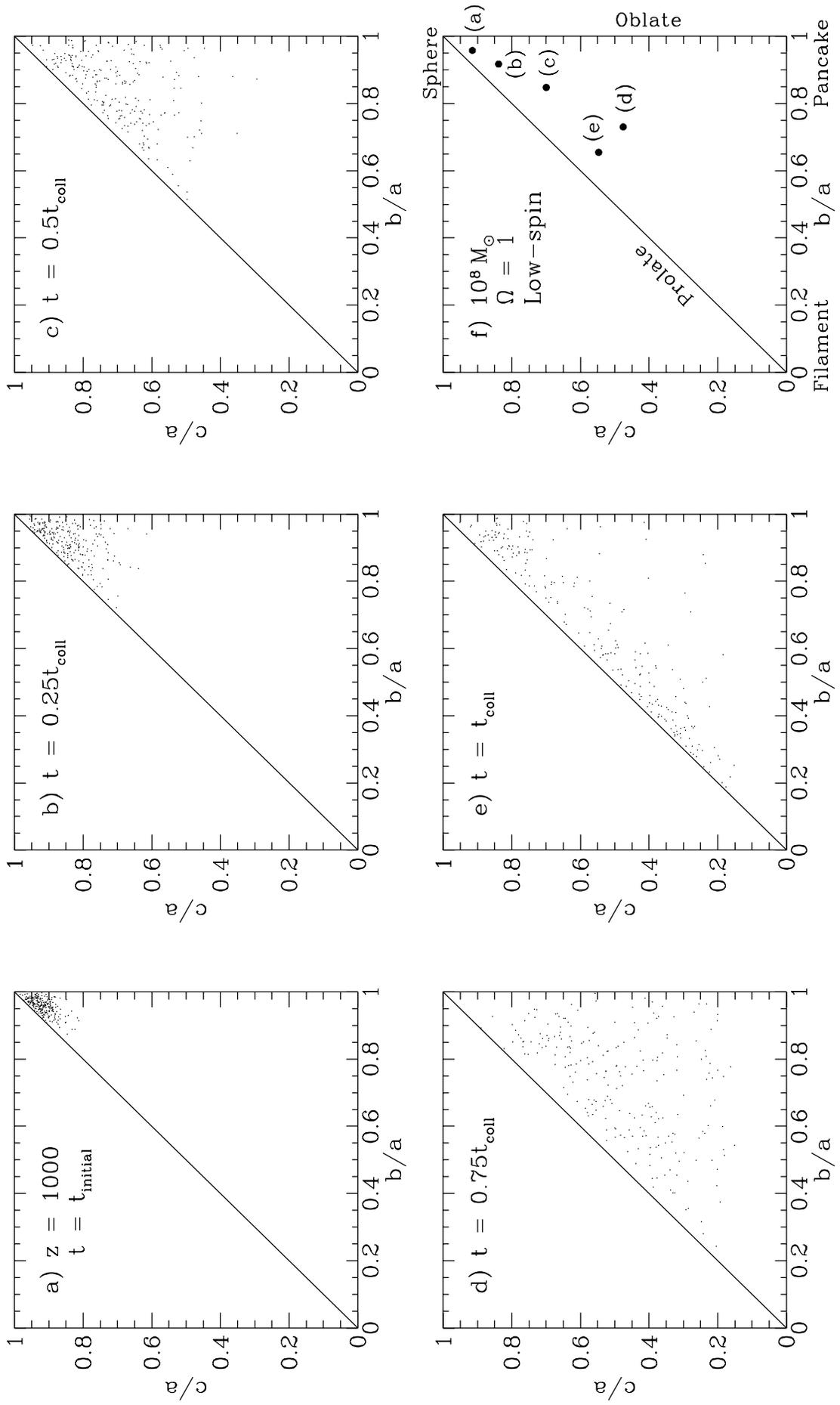
{\bf Fig. 8}: As in Figure 7, but showing only the 232 low-spin objects from
an overall sample of $2\times10^5$ systems.

\noindent
{\bf Fig. 9}: Four individual examples of 
ellipsoids evolved by the model [panel sets (a)-(d)]. The left panel of each set shows 
the lengths of the three semi-axes as functions of time.  The middle panel
of each set shows the overdensity of the ellipsoid (solid line) 
as a function of redshift $z$.  Also shown are the overdensity of a spherical 
top-hat perturbation of equal initial density (dotted line) and the overdensity predicted from
applying the Zel'dovich approximation to the initial potential (dashed line).  
The right panel shows the spin parameter $\lambda$ as a function of time.

\np
\nopagenumbers

$$ \vbox{\tabskip=0pt \offinterlineskip
\def\tablerule{\omit&\multispan4\hrulefill & & \multispan4\hrulefill \cr}
\halign{\strut#& \vrule#\tabskip=1em plus2em& \hfil#\hfil& \hfil#\hfil&
\vrule#& \hfil#&
\vrule#& \hfil#\hfil & \hfil#\hfil & \vrule#\tabskip=0pt\cr
\tablerule
\omit&height3pt& & & & & & & &\cr
& & $\mu_0$ & $P(\mu<\mu_0)$ & & & & $\mu_0$ & $P(\mu<\mu_0)$ & \cr
\omit&height3pt& & & & & & & &\cr
\tablerule
\omit&height3pt& & & & & & & &\cr
& & .20 & .01\% & & & &   1.55 & 50\% & \cr
& & .31 & .1\%  & & & &   1.96 & 75\% & \cr
& & .53 & 1\%   & & & &   2.36 & 90\% & \cr
& & .91 & 10\%  & & & &   3.11 & 99\% & \cr
& & 1.19 & 25\% & & & &   3.69 & 99.9\% & \cr
\omit&height3pt& & & & & & & &\cr
\tablerule }}$$
\medskip

\noindent
{\bf Table 1}: The probability that the most negative eigenvalue $-\mu$ 
of the matrix $S$ in Appendix C is greater than  
a particular threshold $-\mu_0$.

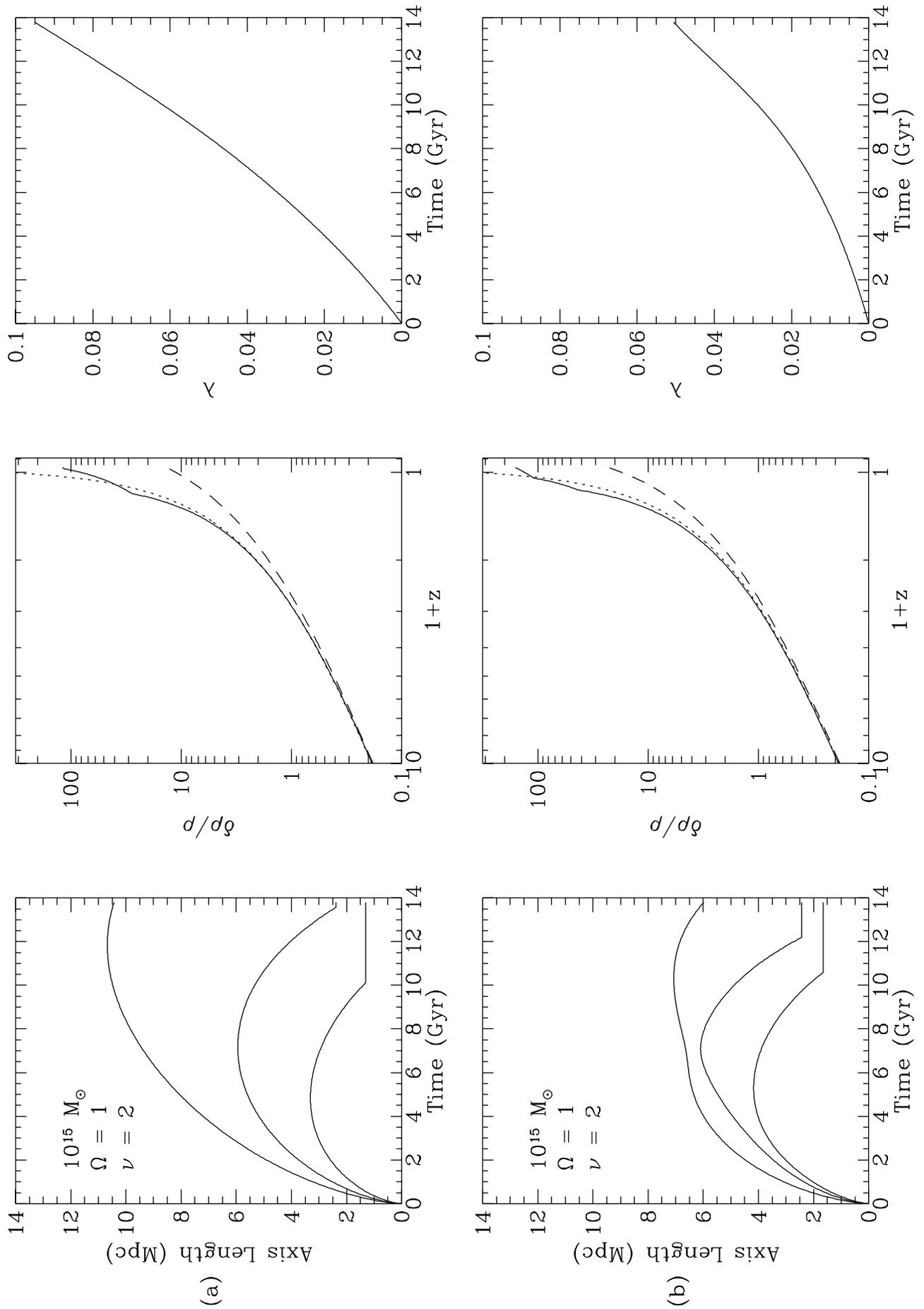
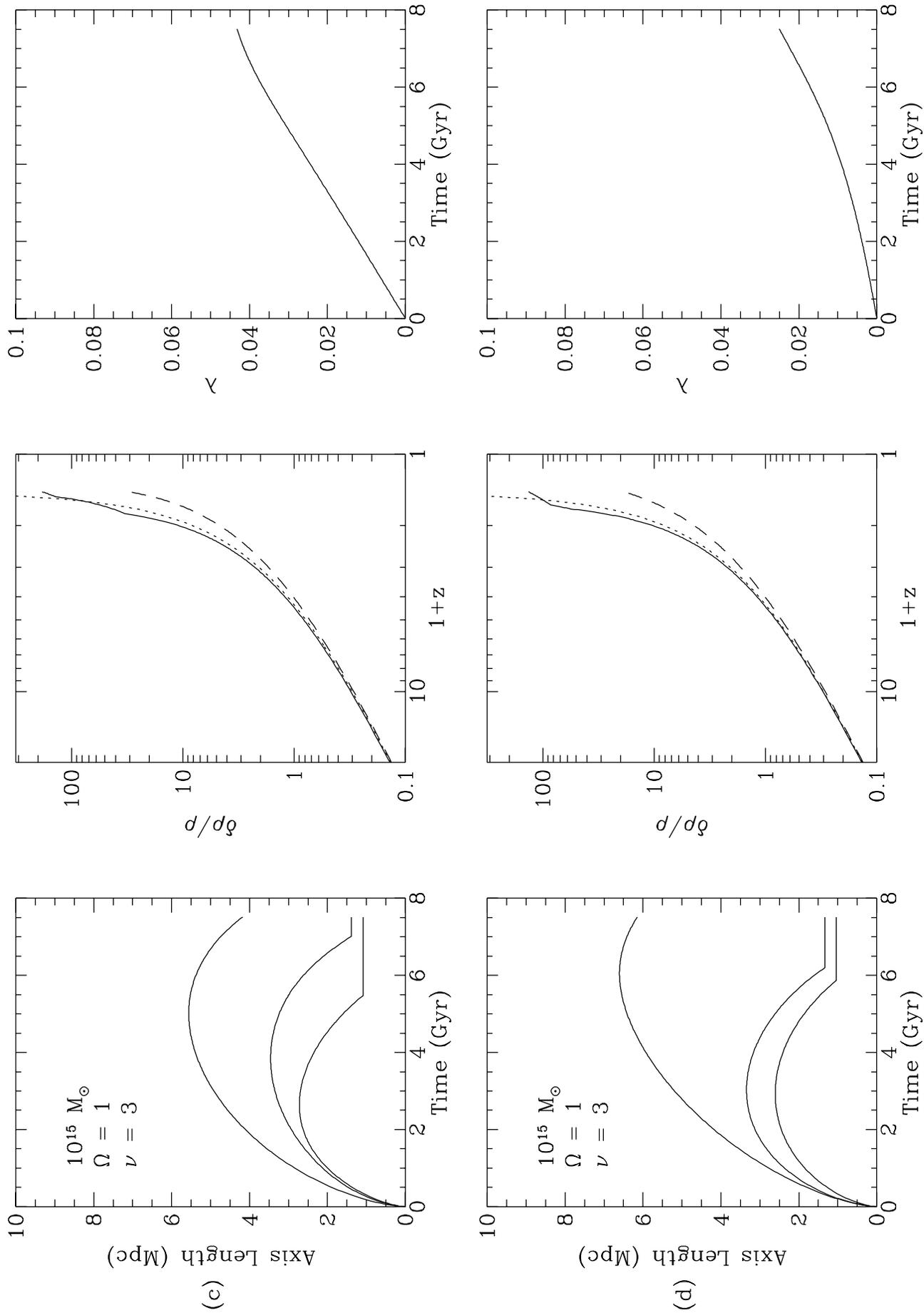
\end